    \pdfoutput=1
   \documentclass[fleqn,usenatbib,usedcolumn]{mnras}
   \usepackage[british]{babel}             
   \usepackage{newtxtext}                  
   \usepackage[slantedGreek]{newtxmath}    
   \let\la=\lesssim 

\usepackage[T1]{fontenc}
\usepackage{ae,aecompl}

 \hypersetup{pdfauthor={T. A. Davis},
               pdftitle={WISDOM Project -- II: Molecular gas measurement of the supermassive black hole mass in NGC4697},
               pdfkeywords={galaxies: individual: NGC 4697 -- galaxies: kinematics and dynamics -- galaxies: nuclei -- galaxies: ISM -- galaxies: elliptical and lenticular, cD},
               bookmarksnumbered=true}
   \volume{{\rm in press}}

\usepackage{amsmath}	
\usepackage{amssymb}	
\usepackage{graphicx,color}
\usepackage{pdflscape}
\usepackage{times,tikz}

\def\hi{\mbox{H\sc{i}}}
\def\kms{km~s$^{-1}$}

\def\msun{M$_{\odot}$}
\def\arcsec{$^{\prime \prime}$}
\definecolor{Mygrey}{gray}{0.75}

\newcommand{\ltsimeq}{\raisebox{-0.6ex}{$\,\stackrel{\raisebox{-.2ex}{$\textstyle <$}}{\sim}\,$}}

\newcommand{\farc}{\mbox{\ensuremath{.\!\!^{\prime\prime}}}}
\mathchardef\mhyphen="2D

\usepackage[compact]{titlesec}
\titlespacing{\section}{0pt}{*2}{*1}

\title[WISDOM: The SMBH in NGC4697]{WISDOM Project -- II: Molecular gas measurement of the supermassive black hole mass in NGC4697} 
\author[Timothy A. Davis et al.]{\parbox{\textwidth}{Timothy A. Davis$^{1}$\thanks{E-mail: \texttt{DavisT@cardiff.ac.uk}}, Martin Bureau$^{2}$,
Kyoko Onishi$^{3,4}$,
Michele Cappellari$^{2}$,
Satoru Iguchi$^{3,4}$,
and Marc Sarzi$^{5}$}
\vspace{0.4cm}\\
\parbox{\textwidth}{$^{1}$School of Physics \&\ Astronomy, Cardiff University, Queens Buildings, The Parade, Cardiff, CF24 3AA, UK\\
$^{2}$Sub-department of Astrophysics, Department of Physics, University of Oxford, Denys Wilkinson Building, Keble Road, Oxford OX1 3RH, UK\\
$^{3}$Department of Astronomical Science, SOKENDAI (The Graduate University of Advanced Studies), Mitaka, Tokyo 181-8588, Japan\\
$^{4}$National Astronomical Observatory of Japan, Mitaka, Tokyo 181-8588, Japan\\
$^{5}$Centre for Astrophysics Research, University of Hertfordshire, Hatfield, Hertfordshire, AL1 9AB, UK}}
\begin{document}
\date{Accepted 2016 December 6. Received 2016 December 6; in original form 2016 October 24}

\pagerange{\pageref{firstpage}--\pageref{lastpage}} \pubyear{2015}

\maketitle

\label{firstpage}

\begin{abstract}
As part of the mm-Wave Interferometric Survey of Dark Object Masses (WISDOM) project, we present an estimate of the mass of the supermassive black hole (SMBH) in the nearby fast-rotating early-type galaxy NGC4697. This estimate is based on Atacama Large Millimeter/submillimeter Array (ALMA) cycle-3 observations of the $^{12}$CO(2--1) emission line with a linear resolution of 29 pc (0\farc53). We find that NGC4697 hosts a small relaxed central molecular gas disc with a mass of 1.6$\times$10$^7$~\msun, co-spatial with the obscuring dust disc visible in optical \textit{Hubble Space Telescope} (\textit{HST}) imaging. We also resolve thermal 1~mm continuum emission from the dust in this disc.
NGC4697 is found to have a very low molecular gas velocity dispersion, $\sigma_{\rm gas}$=1.65$^{+0.68}_{-0.65}$~\kms. This seems to be partially because the giant molecular cloud mass function is not fully sampled, but other mechanisms such as chemical differentiation in a hard radiation field or morphological quenching also seem to be required. We detect a Keplerian increase of the rotation of the molecular gas in the very centre of NGC4697, and use forward modelling of the ALMA data cube in a Bayesian framework with the \textsc{KINematic Molecular Simulation (KinMS)} code to estimate a SMBH mass of (1.3$_{-0.17}^{+0.18}$) $\times$10$^8$ \msun\ and an $i$-band mass-to-light ratio of 2.14$_{-0.05}^{+0.04}$~\msun/L$_{\odot}$ (at the 99\% confidence level). Our estimate of the SMBH mass is entirely consistent with previous measurements from stellar kinematics. This increases confidence in the growing number of SMBH mass estimates being obtained in the ALMA era.
\end{abstract} 

\begin{keywords}
galaxies: individual: NGC 4697 -- galaxies: kinematics and dynamics -- galaxies: nuclei -- galaxies: ISM -- galaxies: elliptical and lenticular, cD 
\end{keywords}

\renewcommand{\thefootnote}{\ensuremath{^\arabic{footnote}}}
\setcounter{footnote}{0}

\begin{figure*} \begin{center}
\begin{tikzpicture}
    \node[anchor=south west,inner sep=0] (image) at (0,0) { 
    	\begin{minipage}[b]{0.52\textwidth}
\begin{tikzpicture}
    \node[anchor=south west,inner sep=0] (image) at (0,0) {\includegraphics[height=9.25cm,angle=0,clip,trim=0cm 0cm 0cm 0.0cm]{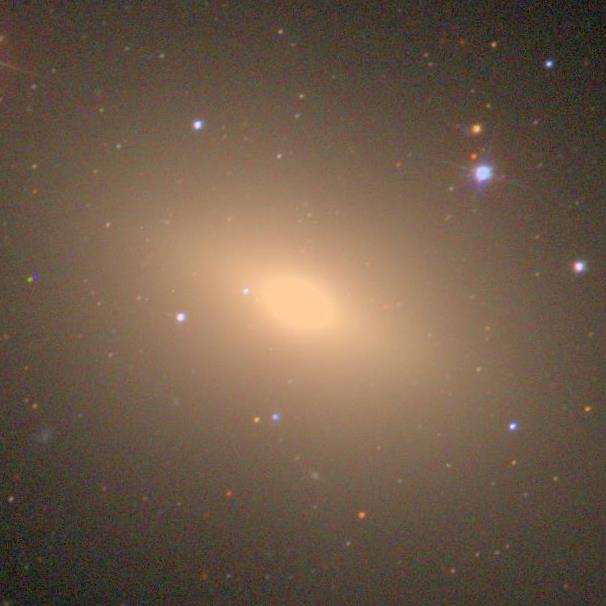}};
    	\begin{scope}[x={(image.south east)},y={(image.north west)}]
        \draw[blue,ultra thick] (0.47,0.47) rectangle (0.53,0.53);
         \node[text=white] at (0.05,0.9) {SDSS};
        	\draw[white,thick] (0.07,0.1) -- (0.22,0.1);	
	\draw[white,thick] (0.07,0.09) -- (0.07,0.11);	
	\draw[white,thick] (0.22,0.09) -- (0.22,0.11);	
	 \node[text=white] at (0.115,0.10) {2 kpc};
           \end{scope}
\end{tikzpicture}\vspace{3.2cm}
	\end{minipage}\hspace{0.5cm}
	\begin{minipage}[b]{0.45\textwidth}
	\begin{tikzpicture}
    \node[anchor=south west,inner sep=0] (image) at (0,0) {\includegraphics[width=8cm,angle=0,clip,trim=0cm 2.9cm 0cm 0.0cm]{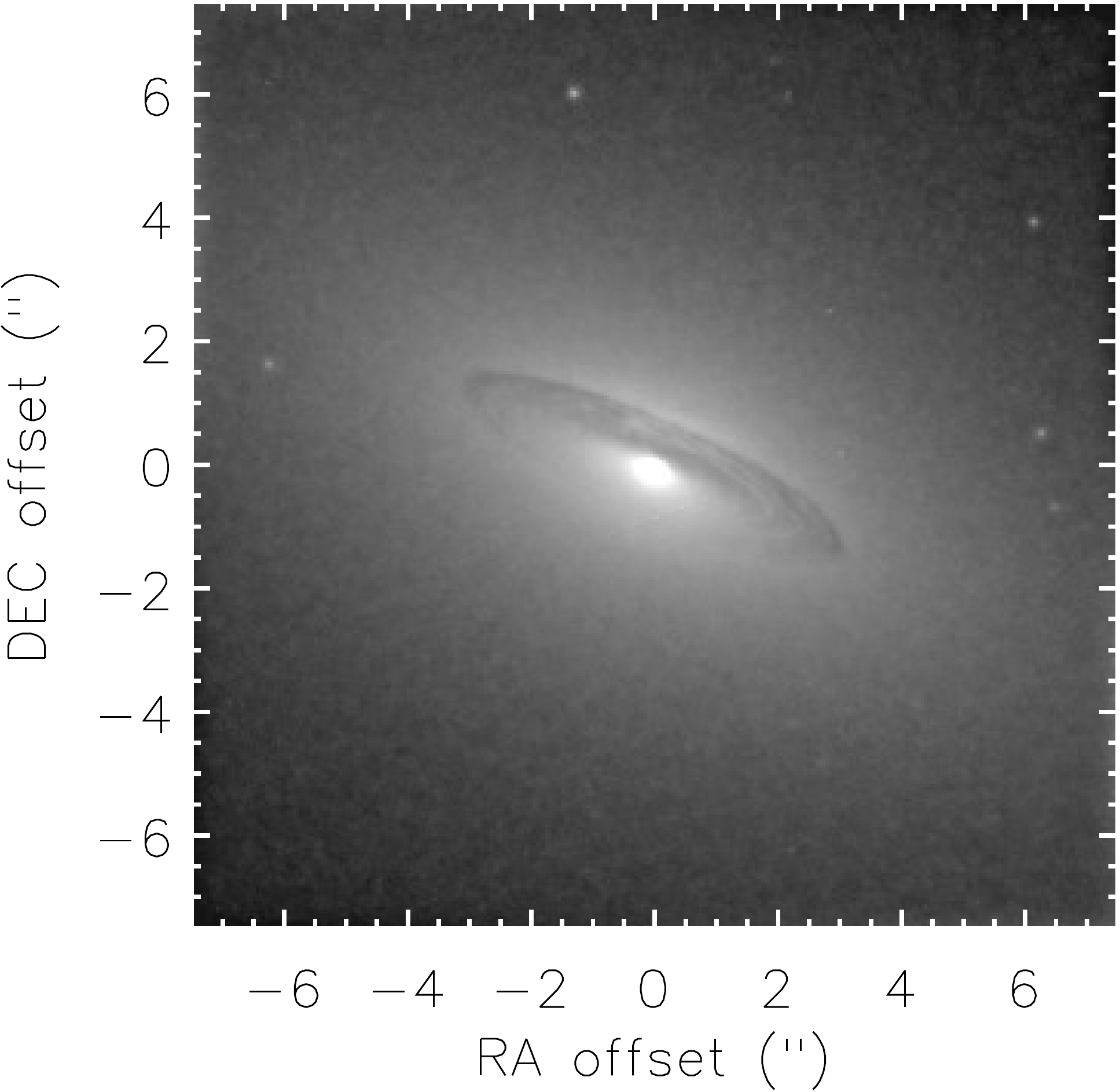}};
     	\begin{scope}[x={(image.south east)},y={(image.north west)}]
 	\draw[white,thick] (0.717,0.1) -- (0.91,0.1);	
	\draw[white,thick] (0.717,0.09) -- (0.717,0.11);	
	\draw[white,thick] (0.91,0.09) -- (0.91,0.11);	
	 \node[text=white] at (0.76,0.1) {200 pc};
           \node[text=white] at (0.23,0.9) {HST};
           \end{scope}
\end{tikzpicture}
	\begin{tikzpicture}
    \node[anchor=south west,inner sep=0] (image) at (0,0) {
    \includegraphics[width=8cm,angle=0,clip,trim=0cm 0cm 0cm 0.0cm]{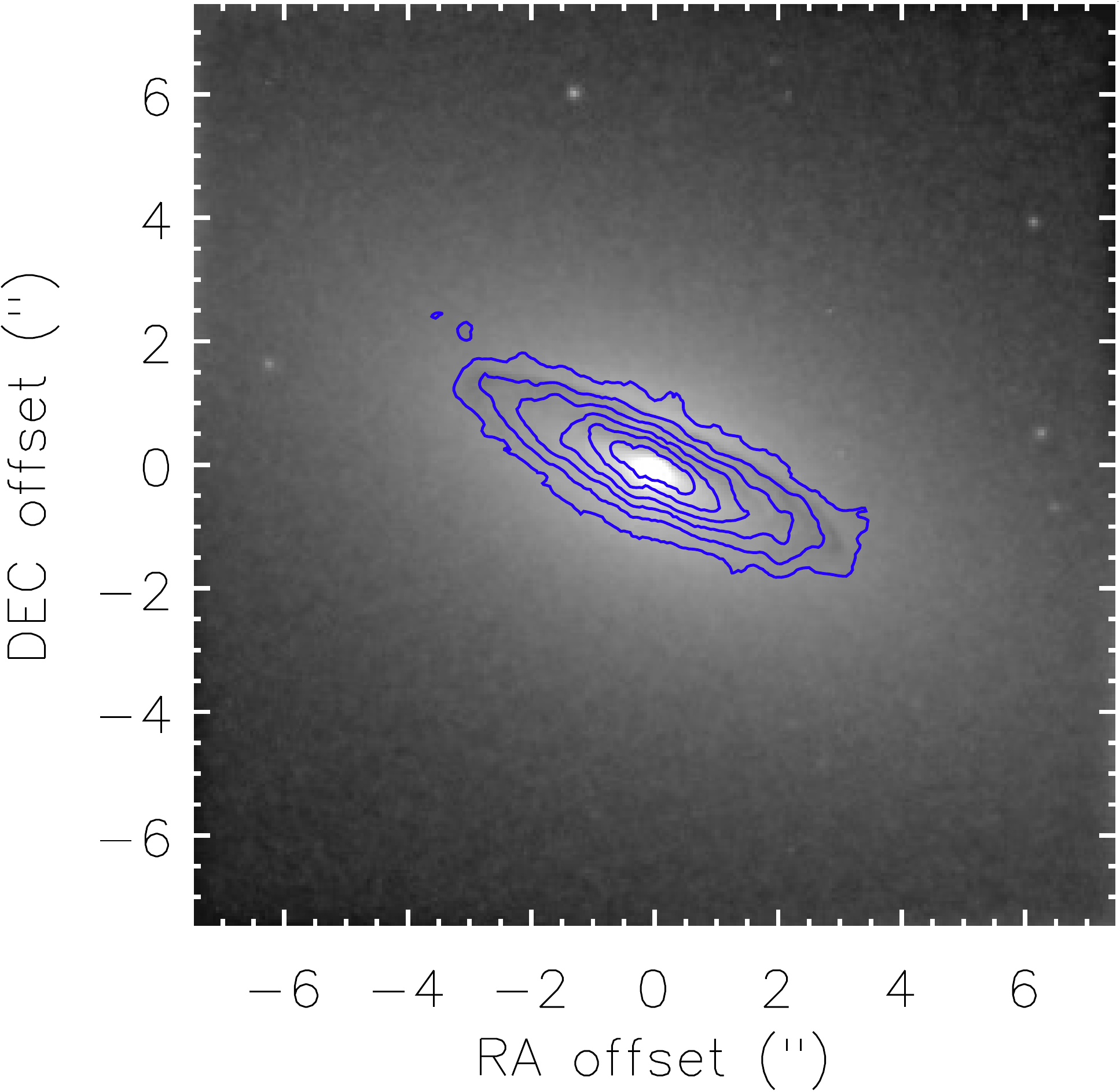}
    };
     	\begin{scope}[x={(image.south east)},y={(image.north west)}]
	\draw[white,thick] (0.717,0.2) -- (0.91,0.2);	
	\draw[white,thick] (0.717,0.19) -- (0.717,0.21);	
	\draw[white,thick] (0.91,0.19) -- (0.91,0.21);	
	 \node[text=white] at (0.76,0.2) {200 pc};
           \node[text=white] at (0.23,0.9) {ALMA CO(2-1)};
           \draw[blue,thick] (0.48,0.915) -- (0.55,0.915);	
           \draw[white,thick] (0.23,0.22) ellipse (0.018 and 0.0173);
           \end{scope}
\end{tikzpicture}
	\end{minipage}
	};
    	\begin{scope}[x={(image.south east)},y={(image.north west)}]
        	\draw[blue,ultra thick] (0.275,0.56) -- (0.63,0.997);	
	\draw[blue,ultra thick] (0.275,0.525) -- (0.63,0.085);
           \end{scope}
\end{tikzpicture}
\caption{\textit{Left panel:} SDSS three-colour ($gri$) image of NGC4697, 4$^\prime$x4$^\prime$ (13.2 kpc $\times$ 13.2 kpc) in size. \textit{Right panel, top:} Unsharp-masked \textit{HST} Advanced Camera for Surveys (ACS) F850LP image of a 825 pc $\times$ 825 pc region (indicated in blue in the left panel) around the nucleus, revealing a clear central dust disc. \textit{Right panel, bottom:} As above, but overlaid with blue $^{12}$CO(2-1) integrated intensity contours from our ALMA observations. The synthesised beam (0\farc54\,$\times$\,0\farc52, 30\,$\times$\,29 pc) is shown as a white ellipse in the bottom-left corner of the panel.}
\label{gal_overview}
 \end{center}
 \end{figure*}
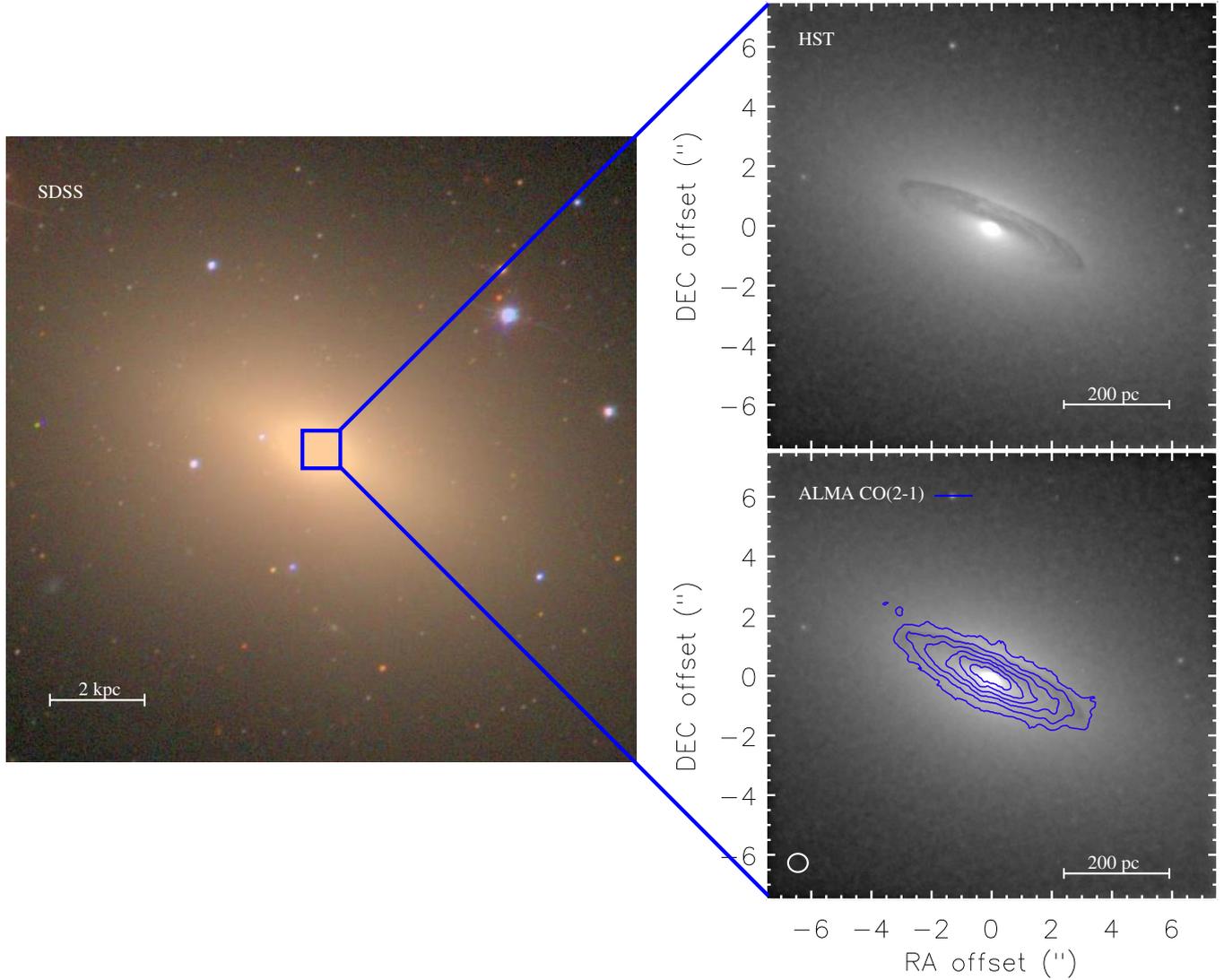

\section{INTRODUCTION}
\noindent Supermassive black holes (SMBHs) are some of the most enigmatic objects in our universe. Evidence has mounted over many years that the majority of massive galaxies have a large SMBH in their centre \citep[see e.g. the review by][]{2013ARA&A..51..511K}.  Although these objects are massive (having typical masses of 10$^6$-10$^{10}$ \msun), their spatial extents are small compared to their host galaxies (with event horizons on AU scales and their gravitational spheres of influence typically $<<$100 pc).
Despite these small sizes, strong correlations between galaxy properties and the masses of the black holes \citep[e.g.][]{1998AJ....115.2285M,2000ApJ...539L..13G,2001ApJ...563L..11G,2009ApJ...698..198G,2013ApJ...764..184M} suggest that SMBHs play a key role in shaping the evolution of massive galaxies.

Theoretical models of galaxy formation are able to reproduce these correlations between galaxies and their SMBHs, and suggest that this co-evolution likely involves self-regulation mechanisms. One possible process that could regulate both SMBH and galaxy growth is feedback from an active galactic nucleus (AGN; \citealt{1998A&A...331L...1S,2008ApJ...676...33D}).
Understanding this co-evolution, and how it changes in galaxies of different masses/morphological types, is vital to understand the impact of SMBHs on galaxy evolution \citep[e.g.][]{2007MNRAS.382.1415S}.

To make progress on this issue observationally, it is crucial to have accurate, well-calibrated ways to measure SMBH masses in a large number of galaxies. A variety of methods exist to estimate the black hole mass for actively accreting objects (e.g. reverberation mapping, \citealt{1998PASP..110..660P}; single epoch line estimates, \citealt{2008ApJ...673..703M}; etc). However, the most reliable method of SMBH mass measurement is usually thought to be spatially-resolved dynamics. In our own Milky Way, this is possible using the resolved motion of individual stars \citep{2008ApJ...689.1044G,2009ApJ...692.1075G}, while for external galaxies measurements can be made using integrated stellar light both with long-slit (e.g. \citealt{1988ApJ...324..701D,1998AJ....115.2285M,2003ApJ...583...92G}) and integral-field spectroscopy (e.g. \citealt{2002MNRAS.335..517V,2009MNRAS.394..660C}). Interstellar gas also provides several complimentary tracers of galaxy potentials. The best of these is maser emission from circumnuclear accretion discs that, when present, allows exquisite SMBH mass estimates \cite[e.g.][]{1995Natur.373..127M,2010ApJ...721...26G}. Unfortunately, suitable systems are quite rare. Ionised gas can also be used to estimate SMBH masses after suitable corrections for asymmetric drift have been applied (e.g. \citealt{1996ApJ...470..444F,2001ApJ...550...65S,2007ApJ...671.1329N}). 

In the last few years substantial improvements in observational capabilities have also allowed molecular gas to be used to estimate SMBH masses. \cite{2013Natur.494..328D} presented the first measurement of this type, estimating the mass of the SMBH in the Virgo cluster fast-rotating early-type galaxy (ETG) NGC4526. \cite{2014MNRAS.443..911D} presented a figure of merit for this technique, and showed that with the Atacama Large Millimeter/submillimeter Array (ALMA) this method should allow SMBH mass measurements in thousands of galaxies across the universe (see also \citealt{2014ApJ...791L..41H} for the prospects of its use in lensed galaxies at very high redshifts). \cite{2015ApJ...806...39O} presented the first use of this technique in a spiral galaxy (NGC1097), while \cite{2016ApJ...822L..28B, 2016ApJ...823...51B} showed its power to resolve the Keplerian increase in velocity around black holes directly. 

As a result, building on some of these small pilot projects (\citealt{2013Natur.494..328D,2015ApJ...806...39O}), we have recently started the mm-{W}ave {I}nterferometric {S}urvey of {D}ark {O}bject {M}asses (WISDOM) project. This project aims to benchmark and test the molecular gas dynamics method, develop tools and best practice, and exploit the growing power of ALMA to better populate and thus constrain SMBH -- galaxy scaling relations. 
The first paper in this series discussed in detail the tools and fitting procedures developed so far, and presented a mass measurement in the nearby fast-rotating early-type galaxy NGC 3665 using CARMA data {(Onishi et al., 2017)}. 

In this work, we present ALMA cycle-3 observations of the molecular gas disc in the centre of the fast-rotating elliptical galaxy NGC4697 (see Fig. \ref{gal_overview}), and use these to estimate the SMBH mass. The SMBH mass in this object is known from previous work using stellar kinematics \citep{2003ApJ...583...92G,2011ApJ...729...21S}, so our new observations allow us to conduct a vital cross check between these two methods.

In Section \ref{target} of this paper we describe our target. In Section \ref{data} we present our ALMA observations, and describe the derived data products. In Section \ref{method} we discuss our dynamical modelling method. In Section \ref{discuss} we discuss our results and compare our SMBH mass measurement with those made by other authors. Finally, we conclude in Section \ref{conclude}. Throughout this paper we assume a distance of 11.4$\pm$1.1 Mpc for NGC4697 \citep{2011MNRAS.413..813C}, as derived from surface brightness fluctuation measurements in \cite{Tonry:2001ei}. At this distance one arcsecond corresponds to a physical scale of 55 pc.

 \section{\uppercase{Target}}
 \label{target}
\noindent  NGC4697 is the brightest galaxy in a poor group (the NGC4697 group) with 5 other lower mass members \citep{2004ApJ...607..810M}. Integral-field observations reveal that despite its optical classification as an E6 elliptical, this object is a fast rotator \citep{2011MNRAS.414..888E}. Figure \ref{gal_overview} shows that it also has a nuclear disc of dust visible in \textit{Hubble Space Telescope} (\textit{HST}) imaging.
NGC4697 has a a total stellar mass of is 1.2$\times$10$^{11}$~\msun, and a luminosity-weighted stellar velocity dispersion within one effective radius of  $\sigma_{\rm e}$= 169 \kms\ \citep{2013MNRAS.432.1709C}.

Both \cite{2003ApJ...583...92G} and \cite{2011ApJ...729...21S} have estimated the SMBH mass in NGC4697, using \textit{HST} long-slit observations to model the stellar kinematics.
They found a black hole mass of $\approx$1.6$\times$10$^8$ \msun. 
There is radio continuum emission from the central regions of this galaxy, leading to its classification as a low-luminosity AGN. The black hole does not seem to be very active, however, with an upper limit on the Eddington ratio from combined radio and X-ray observations  of 10$^{-9}$ \citep{2001ApJ...556..533S,2008ApJ...675.1041W}. 

The star formation rate (SFR) of NGC4697 is not well constrained but is very low. \cite{2014MNRAS.444.3427D} used 22 $\mu$m emission to estimate an upper limit to the SFR of 0.06 \msun\ yr$^{-1}$, and caution that the majority of this flux is likely to arise from the circumstellar envelopes around evolved stars. Assuming all the radio continuum emission in this object comes from star formation sets a more stringent upper-limit of 0.006 \msun\ yr$^{-1}$ \citep{2011ApJ...731L..41B}.
\cite{2013ApJ...770..137F} used \textit{HST} ultraviolet imaging to detect young stellar clusters in this object, and used these to estimate a SFR of 4.6$\times$10$^{-4}$ \msun\ yr$^{-1}$. We note however that \cite{2013ApJ...770..137F} were unable to probe deep inside the dust disc, where the majority of the molecular gas in this system is, so we consider this measurement a lower limit to the total SFR. 

\begin{figure} \begin{center}
\includegraphics[width=0.5\textwidth,angle=0,clip,trim=0cm 0cm 0cm 0.0cm]{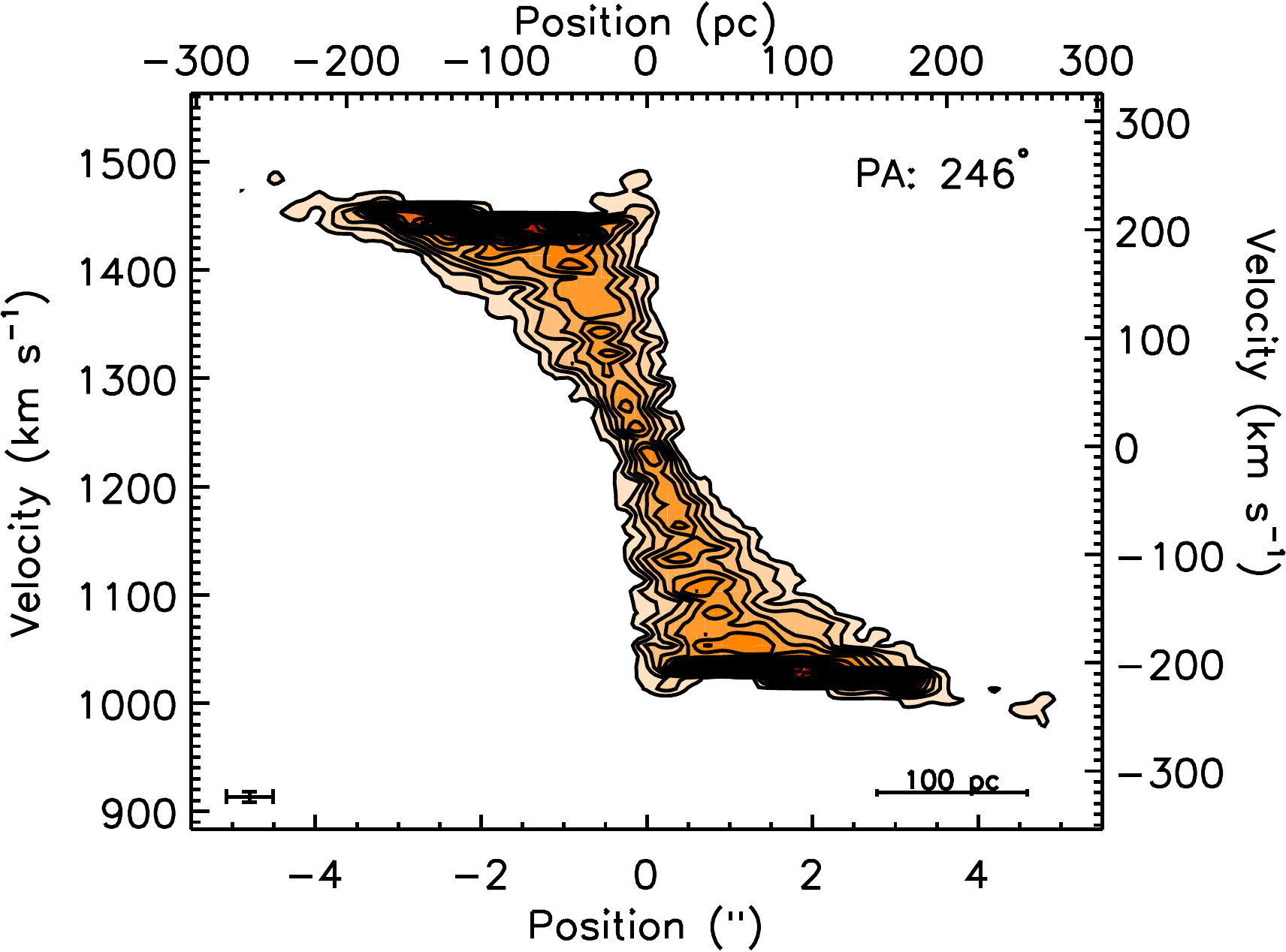}
\caption{Position-velocity diagram of the $^{12}$CO(2-1) emission in NGC4697, extracted along the kinematic major axis. The synthesised beam size (along the major axis) and spectral resolution of our observations is indicated as an error bar in the bottom-left corner. A steep increase of the gas velocity is visible in the galaxy centre, arising from material orbiting around the central SMBH.}
\label{pvdplot}
 \end{center}
 \end{figure}
 
 \section{\uppercase{ALMA data}}
 \label{data}

 The  $^{12}$CO(2--1) line in NGC4697 was observed with ALMA on the 1st of May 2016 as part of the WISDOM project (programme 2015.1.00598.S). The total integration time on source was 574 seconds, observed as a single track. The C36-3 antenna configuration was used, yielding sensitivity to emission on scales up to 17\arcsec. 
An 1850 MHz correlator window was placed over the CO(2--1) line, yielding a continuous velocity coverage of $\approx$2000 \kms\ with a raw velocity resolution of $\approx$1 \kms, sufficient to properly cover and sample the line. Three additional 2 GHz wide low-resolution correlator windows were simultaneously used to detect continuum emission.

The raw ALMA data were calibrated using the standard ALMA pipeline, as provided by the ALMA regional centre staff. Amplitude and bandpass calibration were performed using J1256-0547. The atmospheric phase offsets present in the data were determined using J1246-0730 as a phase calibrator.

We then used the \texttt{Common Astronomy Software Applications} {(\tt CASA)} package to combine and image the resultant visibility file, producing a three-dimensional RA-Dec-velocity data cube (with velocities determined with respect to the rest frequency of the $^{12}$CO(2-1) line). In this work we primarily use data with a channel width of 10 \kms, but in Section \ref{uncertainties} we re-image the calibrated visibilities with a channel width of 3 \kms. In both cases pixels of 0\farc1$\times$0\farc1 were chosen as a compromise between spatial sampling and resolution, resulting in approximately 5 pixels across the beam major axis.  

The data presented here were produced using Briggs weighting with a robust parameter of 0.5, yielding a synthesised beam of 0\farc54$\times$0\farc52 at a position angle of -22$^{\circ}$ (a linear resolution of $\approx$30$\times$29 pc). 
Continuum emission was detected, measured over the full line-free bandwidth, and then subtracted from the data in the $uv$ plane using the {\tt CASA} task {\tt uvcontsub}. The achieved continuum root-mean square (RMS) noise is 35 $\mu$Jy. The continuum-subtracted dirty cubes were cleaned in regions of source emission (identified interactively) to a threshold equal to the RMS noise of the dirty channels.  The clean components were then added back and re-convolved using a Gaussian beam of full-width-at-half-maximum (FWHM) equal to that of the dirty beam.  This produced the final, reduced and fully calibrated $^{12}$CO(2--1) data cubes of NGC4697, with an RMS noise level of 1.1 mJy beam$^{-1}$ in each 10 \kms\ channel (or 2.1 mJy beam$^{-1}$ in each 3 \kms\ channel).

 \subsection{Line emission}

The clean fully calibrated data cube was used to create our final data products. The observed channel maps are presented in Appendix \ref{channelmaps}.
A major-axis position-velocity diagram (PVD; taken along a position angle of 246$^{\circ}$, as determined below, and with a width of 5 pixels) was extracted, and is shown in Figure \ref{pvdplot}.

Zeroth moment (integrated intensity), first moment (mean velocity), and second moment (velocity dispersion) maps of the detected line emission were then created using a masked moment technique. A copy of the clean data cube was first Gaussian-smoothed spatially (with a FWHM equal to that of the synthesised beam), and then Hanning-smoothed in velocity. A three-dimensional mask was then defined by selecting all pixels above a fixed flux threshold of 1.5$\sigma$, adjusted to recover as much flux as possible in the moment maps while minimising the noise.  The moment maps were then created using the un-smoothed cubes within the masked regions only.
The moments are presented in Section \ref{uncertainties}.

We clearly detect a regular disc of molecular gas in NGC4697, with a size of $\approx$390 pc\,$\times$\,100 pc in projection. This gas is regularly rotating and lies coincident with the dust disc visible in \textit{HST} imaging (see Fig. \ref{gal_overview}). A central enhancement of the gas velocity is clearly present around the galaxy centre, that is likely to be the Keplerian increase of the gas velocity around a putative SMBH (see Fig. \ref{pvdplot}). 

The velocity dispersion in this molecular disc seems very low, with clear channelisation present in the upper envelope of the PVD. This suggests the velocity dispersion is small compared with our channel width of 10 \kms. We note that the large line widths visible in Figure \ref{pvdplot} (and in the second moment maps presented in Section \ref{fitting}) are primarily due not to the intrinsic velocity dispersion, but to beam smearing over the velocity gradient at the galaxy centre and the line-of-sight integration through the fairly edge-on disc. This is discussed further in Sections \ref{uncertainties} and \ref{discuss}.

\begin{figure} \begin{center}
\includegraphics[width=0.48\textwidth,angle=0,clip,trim=0cm 0cm 0cm 0.0cm]{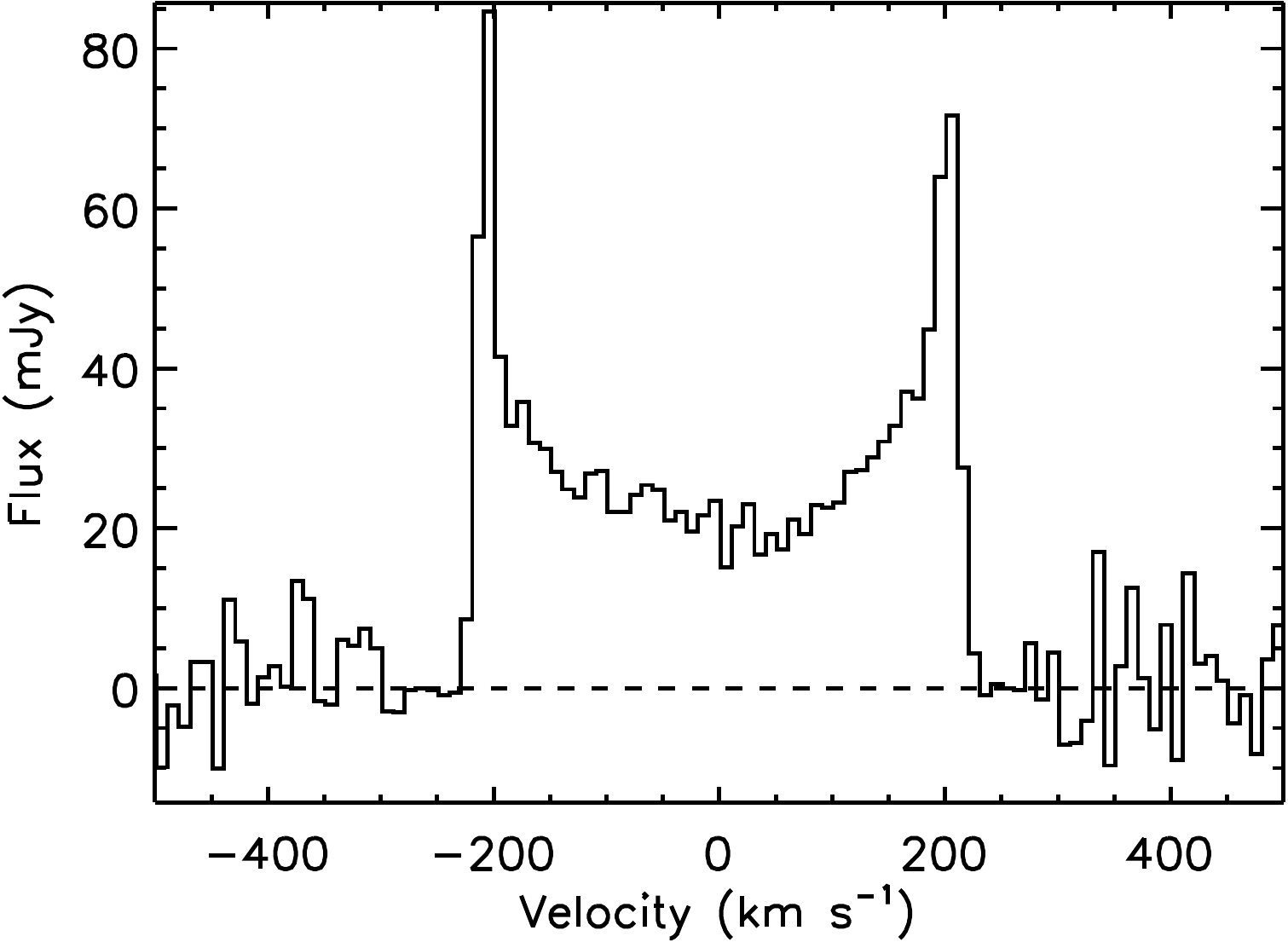}
\caption{Integrated $^{12}$CO(2-1) spectrum extracted from our observed data cube in an 8\arcsec$\times$4\arcsec\ (440 pc $\times$ 220 pc) region around the galaxy centre, covering all the detected emission. A dashed line indicates the zero flux level. The spectrum shows the classic double-horn shape of a rotating disc.}
\label{spectrumplot}
 \end{center}
 \end{figure}

Figure \ref{spectrumplot} shows the integrated $^{12}$CO(2--1) spectrum of NGC4697, with the classic double-horn shape of a rotating disc. The total flux is 13.27~$\pm$~0.07~$\pm$1.3 Jy \kms\ (where the second uncertainty is systematic and accounts for the $\approx$10\% flux calibration uncertainty in the ALMA data). From this measurement, assuming a typical CO(2-1)/CO(1-0) ratio of 0.8 \citep{2008AJ....136.2846B} and a Galactic X$_{\rm CO}$ conversion factor (as in \citealt{2011MNRAS.414..940Y}), we estimate a total H$_2$ mass of (1.62\,$\pm$\,0.01\,$\pm$\,0.36)$\times$10$^7$~\msun. The systematic uncertainty quoted here on the H$_2$ mass includes a contribution from both the flux calibration uncertainty and the uncertainty in our assumed distance. These uncertainties are included as a second error bar where appropriate in the rest of this Section.

Like many ETGs, NGC4697 seems to have a lower SFR per unit gas mass (also known as the star formation efficiency; SFE) than nearby spiral galaxies \citep{2011MNRAS.415...61S,2014MNRAS.444.3427D}.
As described above, the SFR of this system is hard to constrain but likely lies in the range  4.6$\times$10$^{-4}$ \ltsimeq SFR \ltsimeq 0.006 \msun~yr$^{-1}$. 
{The specific star formation rate (star formation rate per unit stellar mass) of NGC4697 is thus very low, between 5$\times$10$^{-14}$ and 4$\times$10$^{-15}$ yr$^{-1}$.}
Using our total H$_2$ mass, we calculate that the SFE is between 2.8$\times$10$^{-11}$ and 3.7$\times$10$^{-10}$~yr$^{-1}$.  This equates to a gas depletion time of between 2.7 and 35 Gyr, lower than in local spirals {(and indeed galaxies of all types with a similar stellar mass)} that typically have depletion times of $\approx$2 Gyr \citep{2011ApJ...730L..13B,2011MNRAS.415...61S}.

\begin{figure*} \begin{center}
\includegraphics[height=6cm,angle=0,clip,trim=0cm 0cm 0cm 0.0cm]{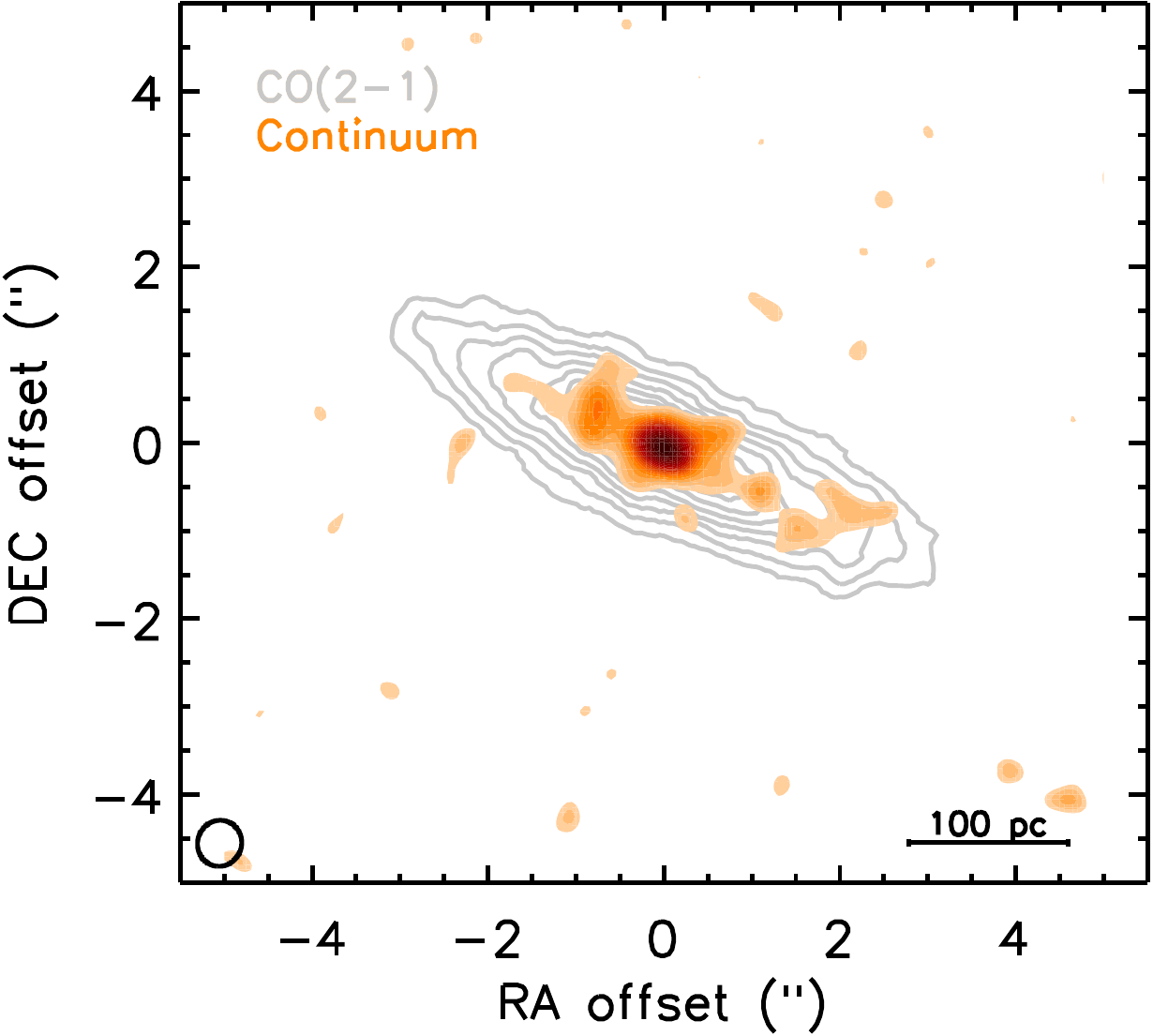}\hspace{0.2cm}
\includegraphics[height=6cm,angle=0,clip,trim=0cm 0cm 0cm 0.0cm]{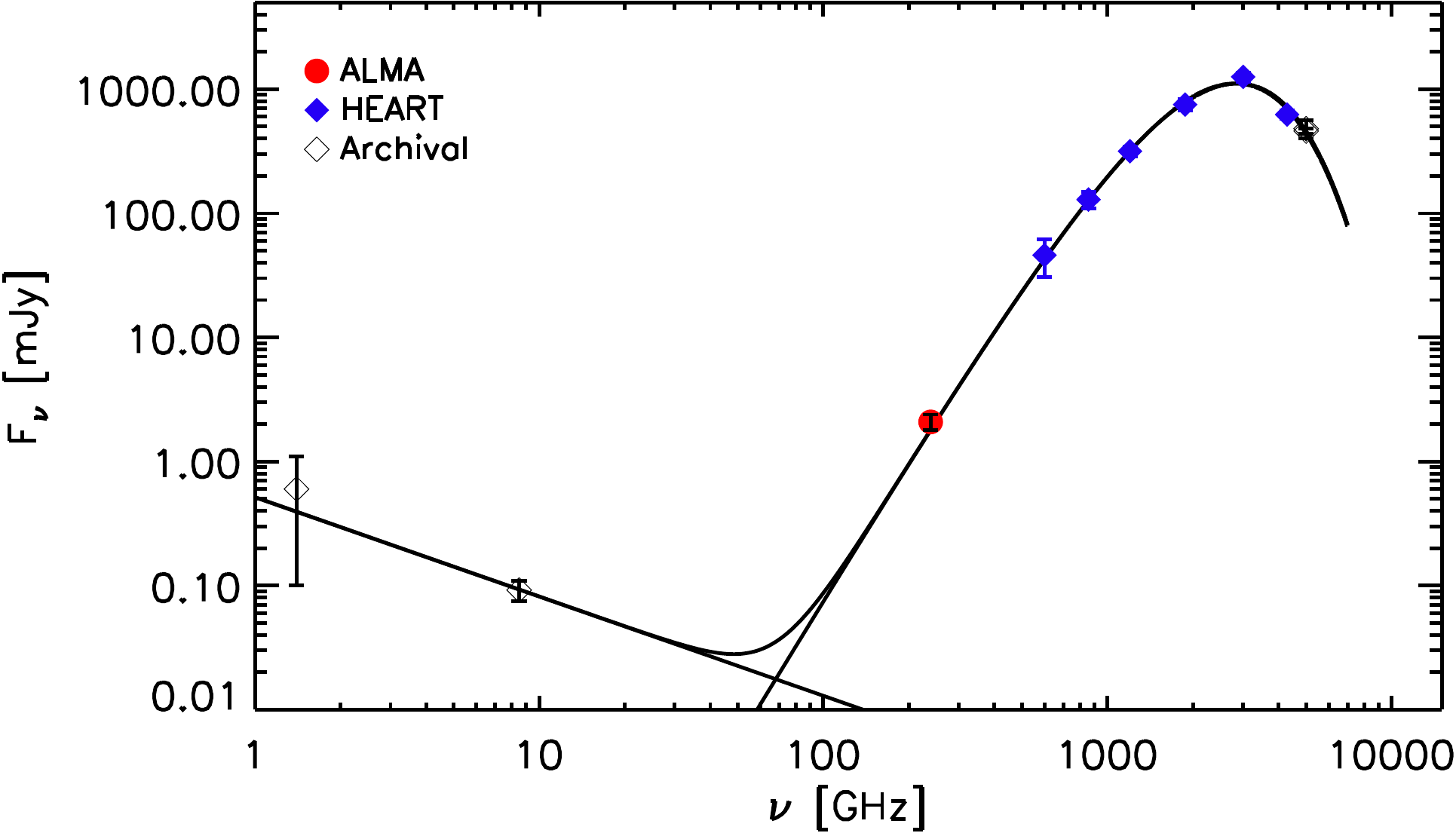}
\caption{\textit{Left panel:} Map of the 1 mm continuum emission from NGC4697 (orange filled contours), {overplotted on the CO(2-1) contours from Figure \protect \ref{gal_overview} (grey)}. \textit{Right panel:} Spectral energy distribution of NGC4697, showing archival radio continuum and mid-infrared observations as black open diamonds, the HEART project reprocessed \textit{Spitzer} and \textit{Herschel} data as blue diamonds, and our total ALMA 1 mm continuum flux measurement as a red circle. The error bar on our ALMA measurement includes the systematic uncertainties as described in the text. The three largely overlapping black lines show the best-fit SED and its infrared dust and radio synchrotron components. The ALMA continuum is consistent with arising entirely from the Rayleigh-Jeans tail of emission from dust.}
\label{contfig}
 \end{center}
 \end{figure*}

 \subsection{Continuum emission}

 As mentioned above, continuum emission was detected in NGC4697. The emission is spatially resolved, as shown in the left panel of Figure \ref{contfig}. A central point source is present, but flux is also detected in an extended structure aligned with the dust disc visible in the \textit{HST} image. Given that the elongation of the continuum emission is very similar to that of the molecular gas disc itself {(as shown by the grey contours in the left panel of Figure \ref{contfig})}, we consider it unlikely that this emission is from a background object. Aperture photometry allows us to estimate a total flux of 2.1\,$\pm$\,0.1\,$\pm$\,0.2 mJy.
 
 Given the evidence for a low luminosity AGN in NGC4697, we wish to determine if this emission is non thermal in origin (i.e. a central AGN point source plus emission from a jet) or if it is consistent with thermal emission from dust. The right panel of Figure \ref{contfig} shows the spectral energy distribution (SED) of this object from 1 GHz to 10 THz. \textit{Herschel} and \textit{Spitzer Space Telescope} data are taken from the Herschel Almanac of Early Types (HEART) project of Smith et al., in prep., while homogenised radio and infrared archival data were taken from the NASA Extragalactic Database\footnote{http://ned.ipac.caltech.edu/} (NED). 

A simple SED model was fitted to these datapoints using the $\chi^2$ minimisation routines {\tt MPFIT} \citep{MarkwardtIDL}. The far-infrared emission was modelled as a modified black body:

\begin{equation}
S_\nu = \frac{\kappa_\nu M_{\rm d} B(\nu,T_{\rm d})}{D^2}\,,
\end{equation}
where $M_{\rm d}$ is the dust mass with dust temperature $T_{\rm d}$, $B(\nu,T_{\rm d})$
is the Planck function, and $D$ is the distance to the galaxy. $\kappa_\nu$
is the dust absorption coefficient, described by a power law
with dust emissivity index $\beta$ such that $\kappa_\nu \propto \nu^\beta$. We here utilise an empirical $\kappa_\nu$, $\kappa_{500\mu m}$ = 0.051 m$^2$ kg$^{-1}$ and $\beta$\,=\,1.8 \citep{2016MNRAS.459.1646C}.

In the radio regime, with only two data points available, we assume the emission arises from synchrotron losses and can thus be represented by a power law. We fix the radio spectral index to $-0.8$ (as typically assumed for synchrotron from supernovae/star formation; \citealt{1992ARA&A..30..575C}). Although not well constrained, allowing the spectral index to vary would drive it to steeper values. We note that radio emission from AGN typically has a flatter spectral index, that would not agree with these data \citep[e.g.][]{2016MNRAS.458.2221N}.

With the parameters described above we fit the HEART and literature data to obtain a best-fit dust mass and temperature (along with the radio power-law normalisation). We find that the data are well fit with M$_{\rm d}$= (2.8\,$\pm$\,0.2\,$\pm$\,1.0)$\times10^5$ \msun\ and  T$_{\rm d}$=  28.7$\pm$0.4 K, which agrees well with the values obtained by Smith et al., in prep. Our ALMA 1 mm continuum measurement, although not included in the fit, lies on the predicted SED. Including the ALMA data in the fit increases the dust mass and decreases the temperature very slightly, but these changes are not significant given our errors. 

Overall it seems that the millimetre continuum we detect is thermal in origin, and comes from the Rayleigh-Jeans tail of the dust emission. The morphology of the continuum source suggests the AGN in this system is contributing to dust heating, but non-thermal emission is not important at this frequency. 

Taking our dust mass estimate and combining it with our H$_2$ gas mass measured above, we derive a molecular gas-to-dust mass ratio of 58\,$\pm$\,7\,$\pm$\,6. This is entirely consistent with a high metallicity in the gas phase, similar to the majority of relaxed ETGs \citep{2012ApJ...748..123S}.

\section{\uppercase{Method \& Results}}
 \label{method}
  \label{results}

 In the above section we detailed the properties of the data we obtained for NGC4697. If we wish to fulfill the goals of this paper, however, we need to ascertain if the data are suitable for estimation of the SMBH mass. The formal sphere of influence (SOI) radius ($R_{\rm SOI}$) of the SMBH of a system is given by
 
 \begin{equation}
R_{\rm SOI} = \frac{GM_{\rm BH}}{\sigma_*^2}\,,
\end{equation}
 where G is the gravitational constant, $M_{\rm BH}$ is the SMBH mass and  $\sigma_*$ is the stellar velocity dispersion. For our target with $M_{\rm BH}$ = 1.6$\times10^8$ \msun\ \citep{2003ApJ...583...92G,2011ApJ...729...21S} and an effective $\sigma_*$ of 169 \kms \citep{2013MNRAS.432.1709C}, $R_{\rm SOI}$= 27.1~pc, similar to the spatial resolution of our observations. However, as \cite{2014MNRAS.443..911D} discuss in detail, for the molecular gas SMBH mass estimation method the formal SOI criterion is not very meaningful. When using a method based on cold gas, the central velocity (and thus mass) profile of the galaxy becomes important.
 
 \cite{2014MNRAS.443..911D} defined a figure of merit ($\Gamma_{\rm FOM}$) for such observations, that takes the above effects into account. $\Gamma_{\rm FOM}$ is equal to unity for a 1$\sigma$ detection of the increased velocity due to a SMBH in the central beam of interferometric observations. For a robust determination of the SMBH mass, \cite{2014MNRAS.443..911D} suggests obtaining data for which $\Gamma_{\rm FOM} > 5$.
 We thus applied the figure of merit criterion, as detailed in Equation 4 of  \cite{2014MNRAS.443..911D}, to our data:
 
 \begin{equation}
\Gamma_{\rm FOM} = \frac{\sqrt{[v_{\rm gal}(r_\theta)^2 - \phi_{\rm BH}(r_\theta)]} - v_{\rm gal}(r_\theta)} { \delta v} \sin{i}\,,\label{base_eqnofmerit}\\
\end{equation}
\noindent where
 \begin{equation}
\delta v = \sqrt{ 0.5({\rm CW})^2 + \delta v|^2_{\rm gal}}\,,
\end{equation}
$r_\theta$ is the radius one synthesised beam width away from the nucleus along the galaxy major axis,  $v_{\rm gal}$ is the velocity predicted from the luminous mass distribution (see below), $\phi_{\rm BH}$ is the gravitational potential of the SMBH, $i$ is the inclination and $\delta v$ is the velocity precision obtainable. Here we chose to include both the error caused by the finite channel width (${\rm CW}$) and that from uncertainties in the mass model ($\delta v|_{\rm gal}$).

For NGC4697 our observational setup implies $\theta$~=~0\farc52 (28.7\,pc) and CW~=~10~\kms.
 $\phi_{\rm BH}$ is calculated with $M_{\rm BH}$~=~1.6$\times10^8$~\msun\ as above. 
Ellipse fitting to the dust disc in the \textit{HST} images yields $i$\,=\,76$^{\circ}$.
In this object v$_{\rm gal}(r_\theta)$\,=\,160~\kms, calculated by multiplying the MGE of \cite{2013MNRAS.432.1894S} with the mass-to-light ratio of \cite{2013MNRAS.432.1709C}, and calculating the circular velocity as described in Appendix A of \cite{2002ApJ...578..787C}. $\delta v|_{\rm gal}$ is assumed to be $\pm$10~\kms (see Section \ref{uncertainties}).
These parameters yield $\Gamma_{\rm FOM}$\,=\,5.9, suggesting we should be able to accurately constrain the SMBH mass in this system.

 In the rest of this Section we therefore outline our method for estimating the SMBH mass and other physical parameters of NGC4697 from the observed molecular gas kinematics.

\begin{table}
\caption{MGE parameterisation of the galaxy light profile}
\begin{center}
\begin{tabular*}{0.4\textwidth}{@{\extracolsep{\fill}}r r r}
\hline
$\log_{10}$ I$_j$ & $\log_{10}$  $\sigma_j$ & $q_j$ \\
$L_{\odot,i}$ pc$^{-2}$ & (") & \\
(1) & (2) & (3)\\
\hline
$^{*}$5.579  &  $^{*}$-1.281  &   $^{*}$0.932\\
      4.783  &  -0.729   &   1.000\\
      4.445   & -0.341   &  0.863\\
      4.239 & -0.006  &   0.726\\
      4.061   &  0.367   &  0.541\\
      3.698  &   0.665  &   0.683\\
      3.314   &  0.809   &  0.318\\
      3.538   &   1.191   &  0.589\\
 \hline
\end{tabular*}
\parbox[t]{0.4\textwidth}{ \textit{Notes:} The central unresolved gaussian, indicated with a star, is removed to minimise the effect of the AGN on our kinematic fitting.}
\end{center}
\label{mgetable}
\end{table}%

\begin{figure*} \begin{center}
\includegraphics[height=6.25cm,angle=0,clip,trim=0cm 0cm 0cm 0.0cm]{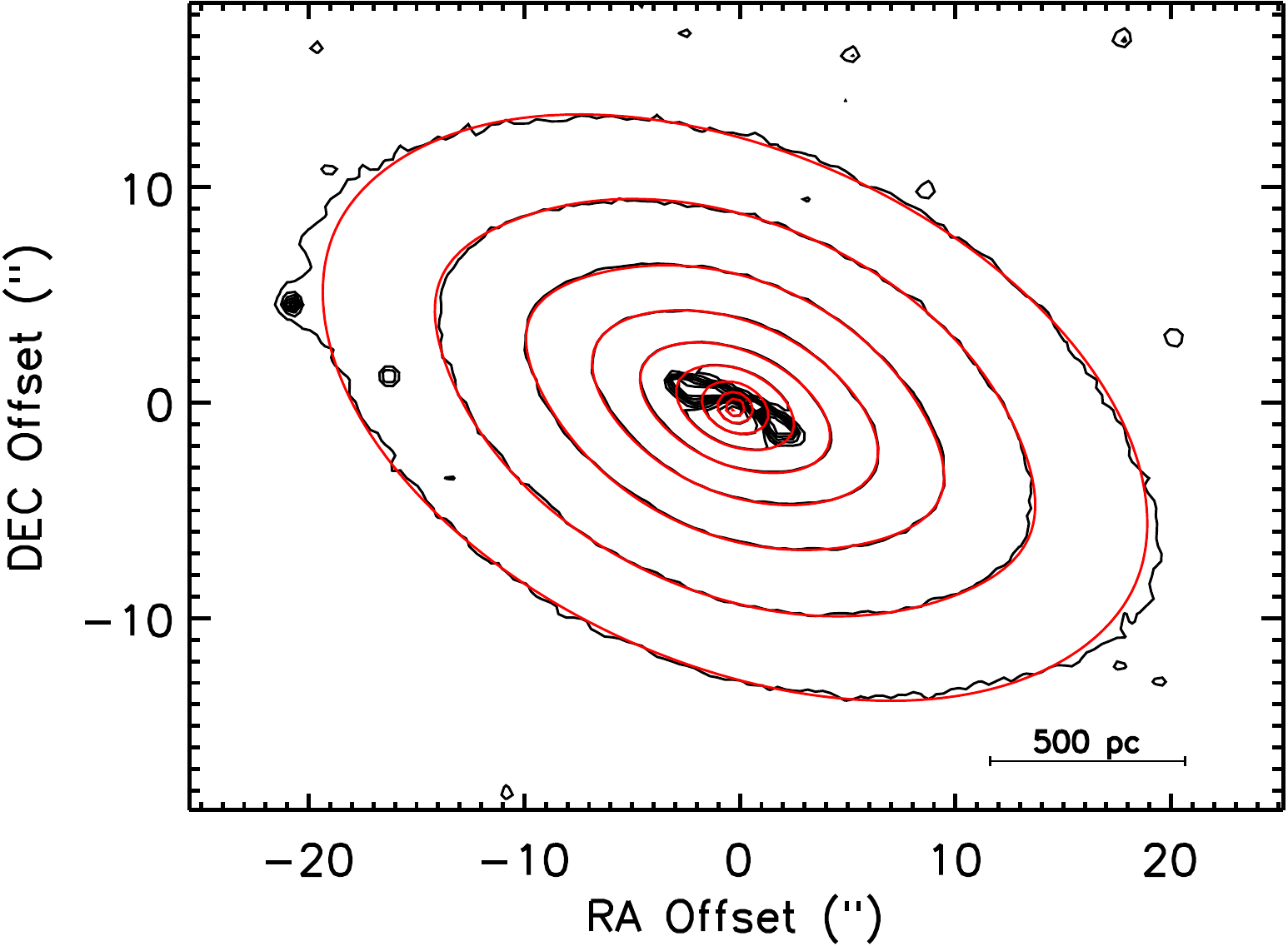}\hspace{0.5cm}
\includegraphics[height=6.25cm,angle=0,clip,trim=0cm 0cm 0cm 0.0cm]{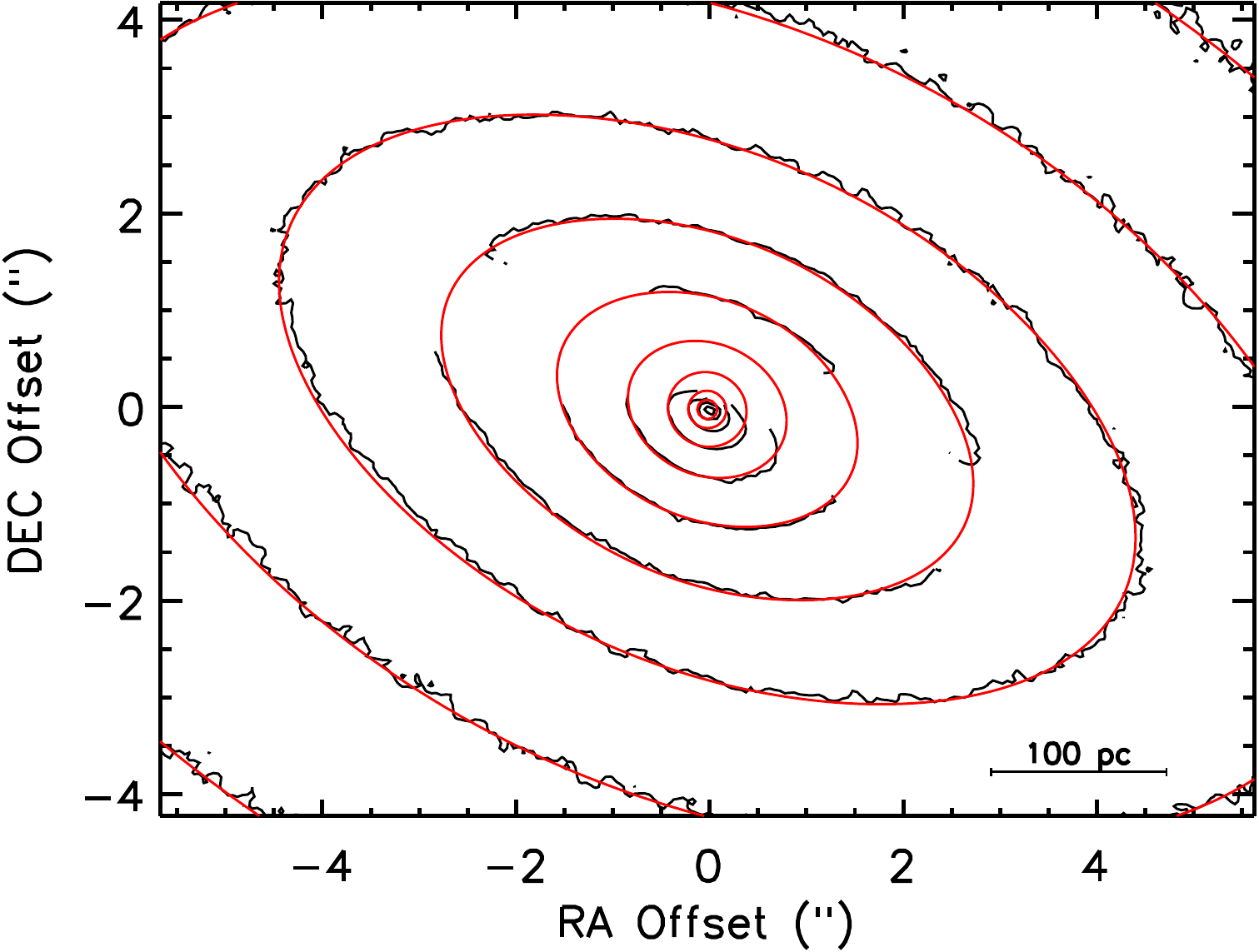}
\caption{Our MGE model of NGC4697 (red contours), overlaid on a \textit{HST} $i$-band (ACS F850LP) image (black contours). The left panel shows the whole galaxy, while the right panel shows a zoom in on the central region. The region masked due to dust is clearly visible to the north of the galaxy nucleus in the left panel, and as breaks in the black contours in the right panel.}
\label{mge_galplot}
 \end{center}
 \end{figure*}

\subsection{Dynamical modelling}

We used a forward modelling approach to estimate the black hole mass (and other physical parameters) of our source. We utilised the publicly available \textsc{KINematic Molecular Simulation} (\textsc{KinMS}\footnote{https://github.com/TimothyADavis/KinMS}) mm-wave observation simulation tool of \cite{2013MNRAS.429..534D}. This tool allows input guesses for the true gas distribution and kinematics and, taking into account the observational effects of beam-smearing, spatial and velocity binning, disc thickness, gas velocity dispersion, etc, produces a simulated data cube that can be compared to observational data. 

To determine the best-fit model parameters and their uncertanties, we used the Markov Chain Monte Carlo (MCMC) code \textsc{KinMS\_mcmc} (Davis et al., in prep.), that couples to the KinMS routines and allows us to fit the data and obtain samples drawn from the Bayesian posterior probability distributions of the fitted parameters. This code fits the entire data cubes produced by interferometers rather than simply the PVD (as discussed in detail in Onishi et al., 2017). 
The simulated cubes use a synthesised beam, pixel size and velocity resolution identical to our observations.

\subsubsection{Gas distribution}
One of the inputs of the \textsc{KinMS} models is an arbitrarily parameterised function that describes the gas surface brightness distribution. At our resolution the molecular gas disc of NGC4697 is remarkably consistent with an exponential disc (that has also been shown to be appropriate for most ETGs; \citealt{2013MNRAS.429..534D}), so this simple form was adopted. In our modelling the exponential disc scale length was left as a free parameter, modified by the \textsc{KinMS\_mcmc} code to obtain a good fit. 
 Various other free parameters of the gas disc are also included in the model. These are the total flux, position angle and inclination of the gas disc, as well as its kinematic centre (in RA, Dec and velocity). We find no evidence of a warp in this galaxy, so the inclination and position angle are each fitted as a single value valid throughout the disc.

\subsubsection{Gas kinematics}

The kinematics of the molecular gas contains contributions from both the luminous stellar mass and the SMBH. 
To remove the contribution of visible matter, and calculate the true mass of any dark object present at the centre of NGC4697, we thus need to construct a luminous mass model. We note that dark matter, while important at larger galactocentric distances, is usually negligible at these radii. Indeed, \cite{2013MNRAS.432.1862C} found that this object has a negligible dark matter fraction within one effective radius, and we are working well within this (at $<$R$_{\rm e}$/7).

We parameterise the luminous matter distribution using a multi-Gaussian expansion (MGE; \citealt{Emsellem:1994p723}) model of the stellar light distribution, constructed using the \textsc{MGE\_FIT\_SECTORS} package\footnote{http:purl.org/cappellari/software} of \cite{2002MNRAS.333..400C}.
Our best-fit MGE model is shown in Figure \ref{mge_galplot} and is tabulated in Table \ref{mgetable}.
 This was constructed from an \textit{HST} Advanced Camera for Surveys (ACS) F850LP image (the longest wavelength available, to minimise dust extinction). 
 
Under the Gaussian density distribution assumption, the model of the stellar light can be de-projected analytically given an inclination (the same as fitted by \textsc{KinMS\_mcmc}). The light model then directly predicts the circular velocity of the gas caused by the luminous matter, via the stellar mass-to-light ratio ($M$/$L$; a free parameter of our model). The $M$/$L$ derived is valid in the F850LP filter band, and is defined in the HST ACS system, although we abbreviate this to $M$/$L_i$ here for brevity. We note that NGC4697 contains an AGN, that likely contributes significantly to the unresolved point source at the galaxy centre. We subtract this point source by removing the innermost (spatially unresolved) Gaussian from our MGE model (as listed in Table \ref{mgetable}). 
We note that including this point source would lower our derived SMBH mass by 0.1 dex, and we include this in our derived uncertainties. In any case this would not significantly alter our results (as discussed in Section \ref{discuss}).

In this work we always assume that the gas is in circular motion, and hence that the gas rotation velocity varies only radially. 
We do, however, also include a free parameter for the internal velocity dispersion of the gas, that is assumed to be spatially constant. 
The effects of deviations from circular motion and of allowing the velocity dispersion to vary radially are discussed in Section \ref{uncertainties}.

\begin{figure*} \begin{center}
\includegraphics[width=\textwidth,angle=0,clip,trim=0cm 0cm 0cm 0.0cm]{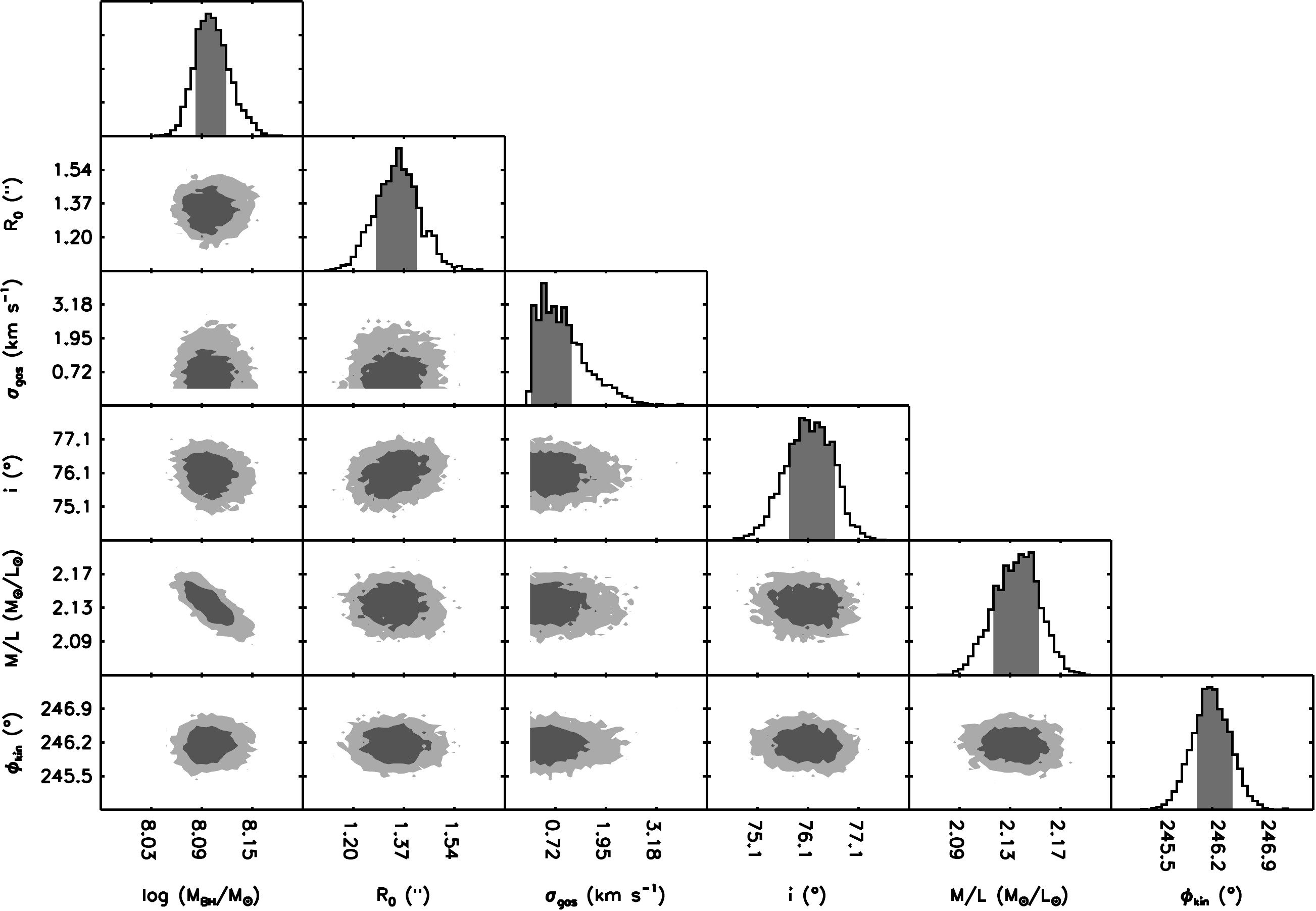}
\caption{
Visualisation of the multidimensional parameter space explored by our fitting procedure for the most important (i.e. physical) fit parameters. In the top panel of each column a one-dimensional histogram shows the marginalised posterior distribution of that given parameter, with the 68\% (1$\sigma$) confidence interval shaded in dark grey. In the panels below, the greyscale shows the two-dimensional marginalisations of those fitted parameters. Regions of parameter space within the 99\% confidence interval are coloured in pale grey, while regions within the 68\% confidence interval are coloured in dark grey. See Table \ref{fittable} for a quantitative description of the likelihoods.}
\label{triangleplot}
 \end{center}
 \end{figure*}

\begin{table*}
\caption{Best fit model parameters and uncertainties.}
\begin{center}
\begin{tabular*}{0.8\textwidth}{@{\extracolsep{\fill}}l r c r r r r}
\hline
Parameter & \multicolumn{3}{c}{Search range} & Best fit & Error (68\% conf.) & Error (99\% conf.)\\
(1) &  \multicolumn{3}{c}{(2)} & (3) & (4) & (5)\\
\hline
\\
Black hole:&&&&&&\\\hline
SMBH mass (10$^8$ \msun) & 0.0006 &$\rightarrow$\hspace{-2cm}& 63.0 & 1.26 & - 0.06, +0.03 & - 0.14, +0.18\\
Stellar $M$/$L$ (\msun/$L_{\odot,i}$) & 0.1 &$\rightarrow$\hspace{-2cm}& 4.0 & 2.14 & $\pm$0.02 & - 0.05, +0.04\\
\\
Molecular gas disc: & & & & & & \\\hline
Position angle ($^\circ$) & 0.0 &$\rightarrow$\hspace{-1cm}& 359.9 & 246.2 & - 0.2, +0.3 & $\pm$0.7\\
Inclination ($^\circ$) & 72.0 &$\rightarrow$& 89.0 & 76.1 & - 0.4, +0.5 & $\pm$1.1\\
Scale length (") & 0.0 &$\rightarrow$& 10.0 & 1.35 & - 0.06, +0.05 & - 0.18, +0.21\\
Velocity dispersion (km s$^{-1}$)$^{*}$ & 0.1 &$\rightarrow$& 30.0 & 0.82 & - 0.63, +0.23& - 0.76, +1.83\\
\\
Nuisance parameters: & & & & & & \\\hline
Luminosity scaling & 1.0 &$\rightarrow$& 50.0 & 17.1 & - 0.8, +0.7 & - 1.9, +2.0\\
Centre X offset (") & -1.0 &$\rightarrow$& 1.0 & -0.10 & $\pm$0.01 & - 0.04, +0.03\\
Centre Y offset (") & -1.0 &$\rightarrow$&1.0 & -0.03 & $\pm$0.01 & - 0.03, +0.02\\
Centre velocity offset (km s$^{-1}$) & -20.0 &$\rightarrow$& 20.0 & 0.15 & - 0.43, +0.35 & - 1.22, +1.25\\
 \hline
\end{tabular*}
\parbox[t]{0.8\textwidth}{ \textit{Notes:} Column 1 lists the fitted model parameters, while Column 2 lists the associated prior. Priors are uniform in linear space (or in logarithmic space for the SMBH mass only). The posterior distribution of each parameter is quantified in the third to fifth columns (see also Fig. 3). The X, Y and velocity offset nuisance parameters are defined relative to the ALMA data phase centre position (RA=12:38:35.91, DEC=-05:48:03.1, V=1241 \kms). A star ($^{*}$) indicates that this parameter is not well constrained by our data with 10 \kms\ channels, as discussed in detail in the text.}
\end{center}
\label{fittable}
\end{table*}%

\subsection{Bayesian analysis}
\label{fitting}

As mentioned above, we use a Bayesian analysis technique to identify the best set of model parameters from our data cube. This allows us to obtain samples drawn from the posterior distribution of the ten model parameters (see Table \ref{fittable}). A full description of this MCMC code will be published in Davis et al., in prep. In brief, we utilize an MCMC method with Gibbs sampling and adaptive stepping to explore the parameter space. This code has been cross checked using a newly developed version of the \textsc{KinMS} routines in the \textsc{python} language (denoted \textsc{KinMSpy}) coupled with the well tested MCMC code \textsc{emcee} \citep{2013PASP..125..306F}. These tests showed that our algorithm performs well, and finds the same best-fit values and parameter ranges.

\subsubsection{Covariance matrix and likelihood} 

As our data are approximately Nyquist sampled spatially, the synthesised beam induces strong correlations between neighbouring spaxels in the data cube. There are two possible approaches to deal with this issue: including the full covariance matrix when calculating the likelihood, or working in the $UV$-plane where the datapoints are uncorrelated. In this work we take the first approach, while the latter will be explored in a future work.

We are able to calculate the full covariance matrix analytically\footnote{following the scheme outlined at http://web.ipac.caltech.edu/staff/fmasci/ home/astro\_refs/PixelNoiseCorrelation.pdf} as we know the synthesised beam shape. We summarise the basic method used below. In what follows, we assume that adjacent velocity channels are independent, and so the same two-dimensional covariance matrix can be used in each channel. Some smoothing is applied to raw velocity channels by the ALMA pipeline, but our assumption of independence is good as the channel width of the processed data is more than twice the raw instrumental channel width.

We assume that each channel of our data cube is formed from some `true' image ($I$) where all pixels are uncorrelated, i.e. the flux in each pixel is a random variable independent of all other fluxes in the image. 
Any channel from this cube that is $N_{\rm x}$ by $N_{\rm y}$ pixels in size contains $N_{\rm x} N_{\rm y}$ pixels, denoted $P_{ij}$, where $i$ runs from 1 to $N_{\rm x}$ and $j$ from 1 to $N_{\rm y}$. This pixel matrix can be represented as a vector composed of elements $p_n$, where $n=1$ to $N_{\rm t}\,\equiv\,N_{\rm x} N_{\rm y}$, where

\begin{eqnarray}
p_{ i + (j -1) N_{\rm x}} = P_{ij},\label{vectoriseme}
\end{eqnarray}

This `true' image has been observed with a response function ($B$) that is oversampled by the pixel size. 
In our case $B$ is known to be isoplanatic (as the ALMA beam, to first order, is spatially invariant), and the observed data $O$ thus arises from the discrete convolution

\begin{eqnarray}
\label{convoleq}
O = I \ast B.
\end{eqnarray}

Let us define the ALMA beam as an elliptical gaussian with a fixed size and position angle ($\mathfrak{B}$). Within a channel of size $N_{\rm x}$ by $N_{\rm y}$ (as above), the beam $\mathfrak{B}_{ij}$ can be vectorised to $\mathfrak{B}_{n}$ using Equation \ref{vectoriseme}. This beam can be centred on any point $p$ within the channel. In vectorised form we denote this $\mathfrak{b}_n(p)$. We use this beam as the response function, and map it to a response matrix $R_{mn}$, which is  $N_{\rm x}N_{\rm y} \times N_{\rm x}N_{\rm y}$ in size. The columns of this matrix can be built up as

\begin{eqnarray}
R_{0m} = \mathfrak{b}_n(p=0), \\
R_{1m} = \mathfrak{b}_n(p=1),\nonumber \\
...\nonumber
\end{eqnarray}

The resulting response matrix is both sparse and diagonally banded. The covariance matrix ($\mathfrak{C}$) can then be calculated from this response matrix as

\begin{eqnarray}
\mathfrak{C} = R^T R\,\sigma^2,
\end{eqnarray}
where $\sigma$ is the RMS noise of our data cube, estimated in central line free regions of the cube. We have here assumed that the observational error in each pixel is the same (i.e. the RMS noise in the data cube does not vary spatially). This is a reasonable assumption in the central region of the ALMA primary beam.

 The resulting covariance matrix is large (up to 4096$\times$4096 pixels for the fitting areas used in this paper), and it has a large condition number. As such, we do not invert it directly to calculate the likelihood, but instead introduce a modified Cholesky (LDL$^T$) factorisation step \citep{1981prop.book.....G} to avoid loss of numerical precision when calculating the inverse ($\mathfrak{C}^{-1}$).

We use a standard logarithmic likelihood function based on the $\chi^2$ distribution, calculated by comparing the CO flux in each pixel of the three-dimensional data cube $D_{xyv}$ (which has size $N_x$ $\times$ $N_y$ in the spatial direction and $N_v$ in the spectral direction) with that in the model $M_{xyv}$. 
The inverse covariance matrix, as described above, is used to compute the log-likelihood ($\mathcal{L}$) in the standard way (see e.g. Eqn. 18 in \citealt{2016arXiv160708538C}).

\begin{eqnarray}
\delta_{xyv}\equiv D_{xyv}-M_{xyv},\\
\chi^2 = \sum_{i=0}^{N_v} \delta_{xyi}^T \,\mathfrak{C}^{-1}\, \delta_{xyi},\\
\mathcal{L} = -\frac{1}{2} (\chi^2 - N_xN_yN_v).
\end{eqnarray}
 It is this likelihood that we minimise in our fitting procedure.

\subsubsection{Fitting process}

To ensure our kinematic fitting process converges, we set reasonable priors on some of the parameters. These are listed in Table \ref{fittable}. The kinematic centre of the galaxy was constrained to lie within two beam-widths of the optical galaxy centre position. The systemic velocity was allowed to vary by $\pm$20 \kms\ from that found by optical analyses. The gas velocity dispersion was constrained to be less than 30 \kms\ and the disc scale length was constrained to be less than 10\arcsec. The $M$/$L$ was allowed to vary between 0.1 and 4.0 \msun/L$_{\odot,i}$. The inclination of the gas disc was allowed to vary over the full physical range allowed by the MGE model. Good fits were always found well within these ranges. A flat prior was used for each of these parameters (an assumption of maximal ignorance). The prior on the SMBH mass was flat in log-space, with the mass allowed to vary between log$_{10}(\frac{M_{\rm BH}}{\mathrm{M}_{\odot}})$~=~4.8 and 9.8.

Once the MCMC chains converged, we ran the best chain 10$^5$ steps (with a 10\% burn-in) to produce our final posterior probability distribution.  For each model parameter these probability surfaces were then marginalised over to produce a best-fit value (median of the marginalised posterior distribution) and associated 68 and 99\% confidence levels (CL). These are also listed in Table \ref{fittable}. Figure \ref{triangleplot} shows the one- and two-dimensional marginalisations of the physical galaxy parameters (i.e. not the kinematic centre and luminosity scaling, that are nuisance parameters). We note that the gas velocity dispersion is very low and not well constrained by our data, hence the clipping at the low end of our prior visible in those panels. This is discussed in more detail in Section \ref{uncertainties}.

\begin{figure*} \begin{center}
\begin{tikzpicture}
    \node[anchor=south west,inner sep=0] (image) at (0,0) {   
\includegraphics[height=6cm,angle=0,clip,trim=0cm 1.5cm 0cm 0.0cm]{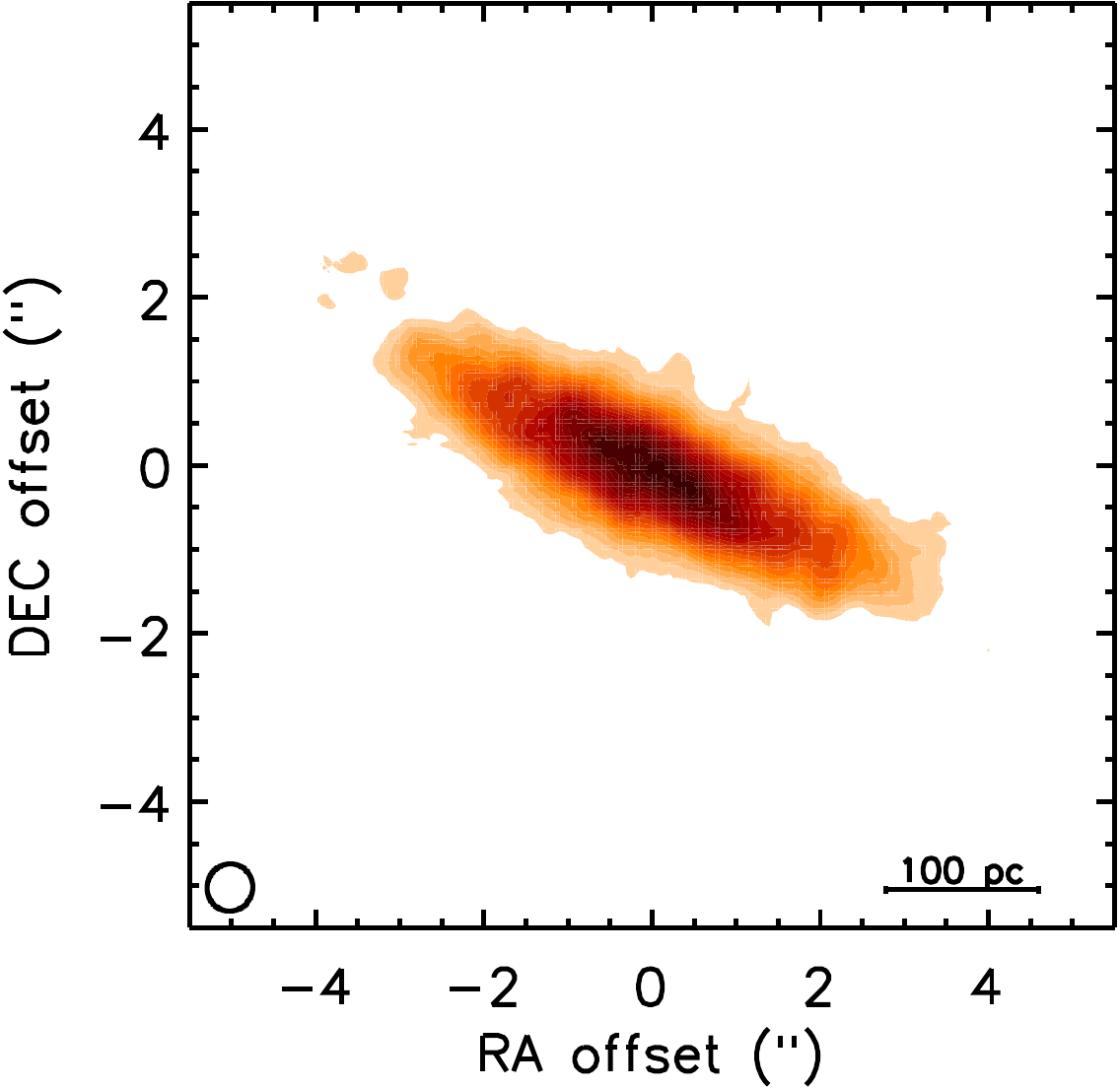}
\includegraphics[height=6cm,angle=0,clip,trim=1.75cm 1.5cm -3.6cm 0.0cm]{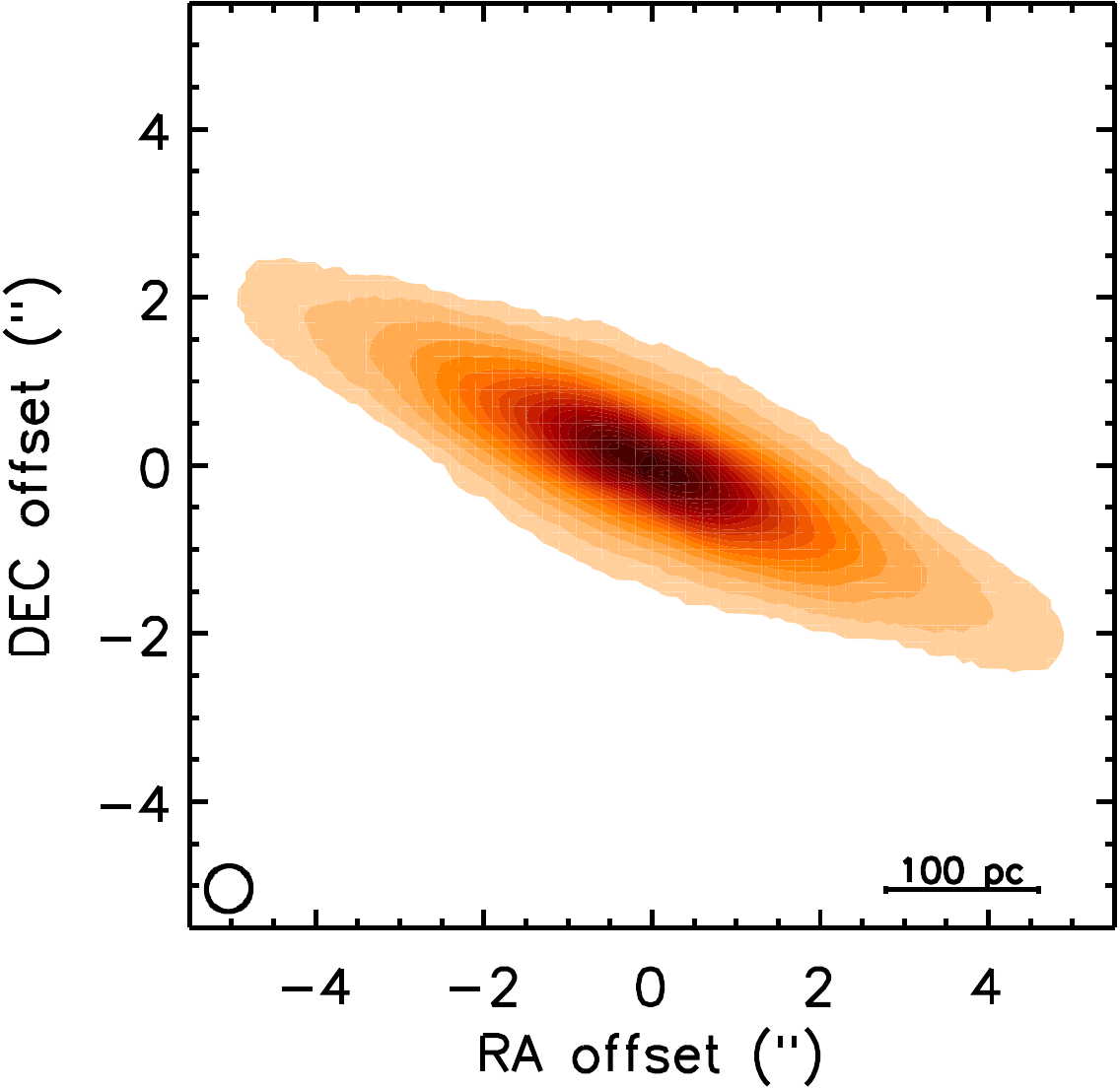}};
    	\begin{scope}[x={(image.south east)},y={(image.north west)}]
          \node[text=black] at (0.125,0.9) {Data};
          \node[text=black] at (0.525,0.9) {Model};
           \end{scope}
\end{tikzpicture}\\
\includegraphics[height=6cm,angle=0,clip,trim=0cm 1.5cm 3.3cm 0.0cm]{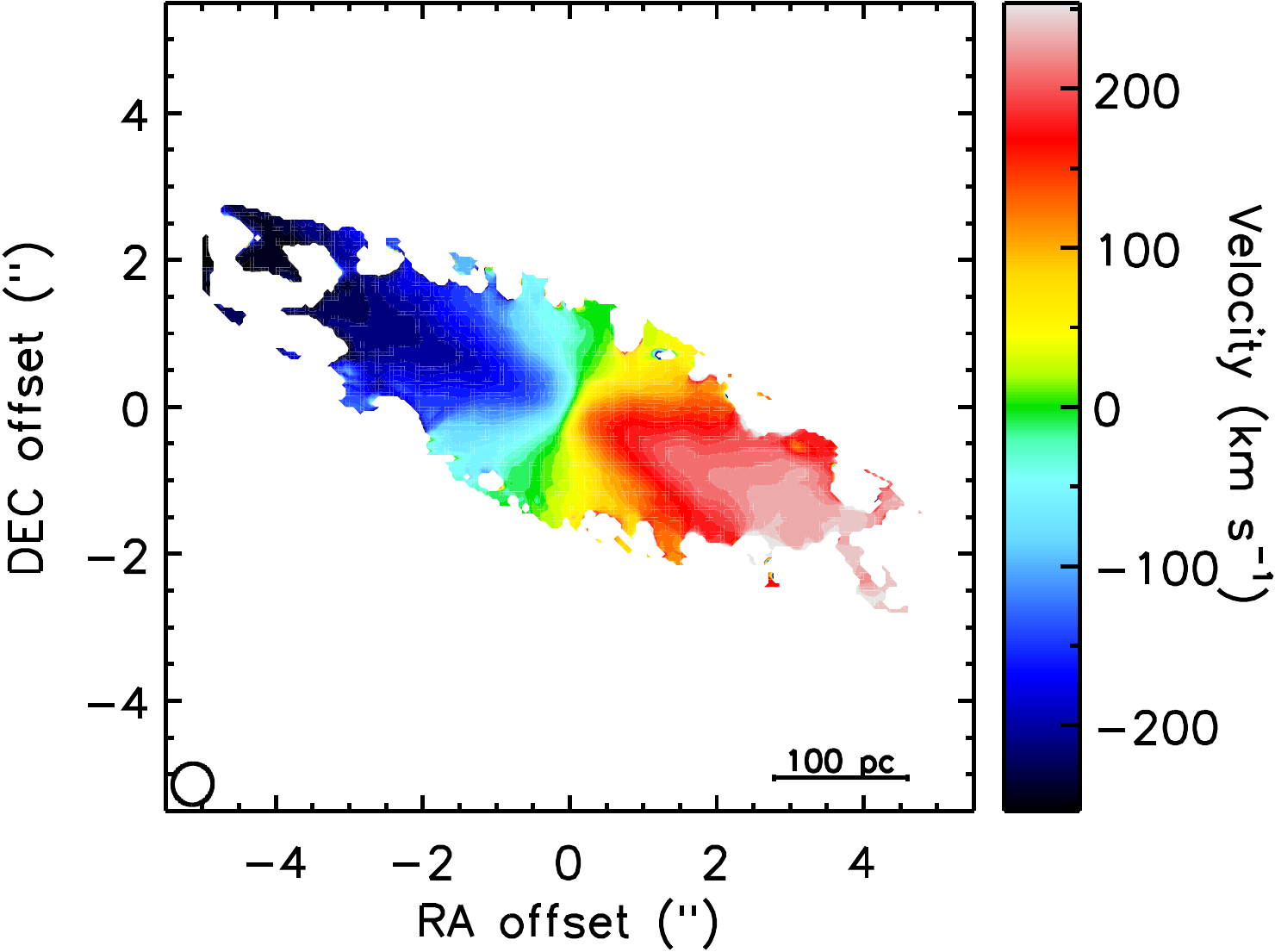}
\includegraphics[height=6cm,angle=0,clip,trim=1.9cm 1.5cm 0cm 0.0cm]{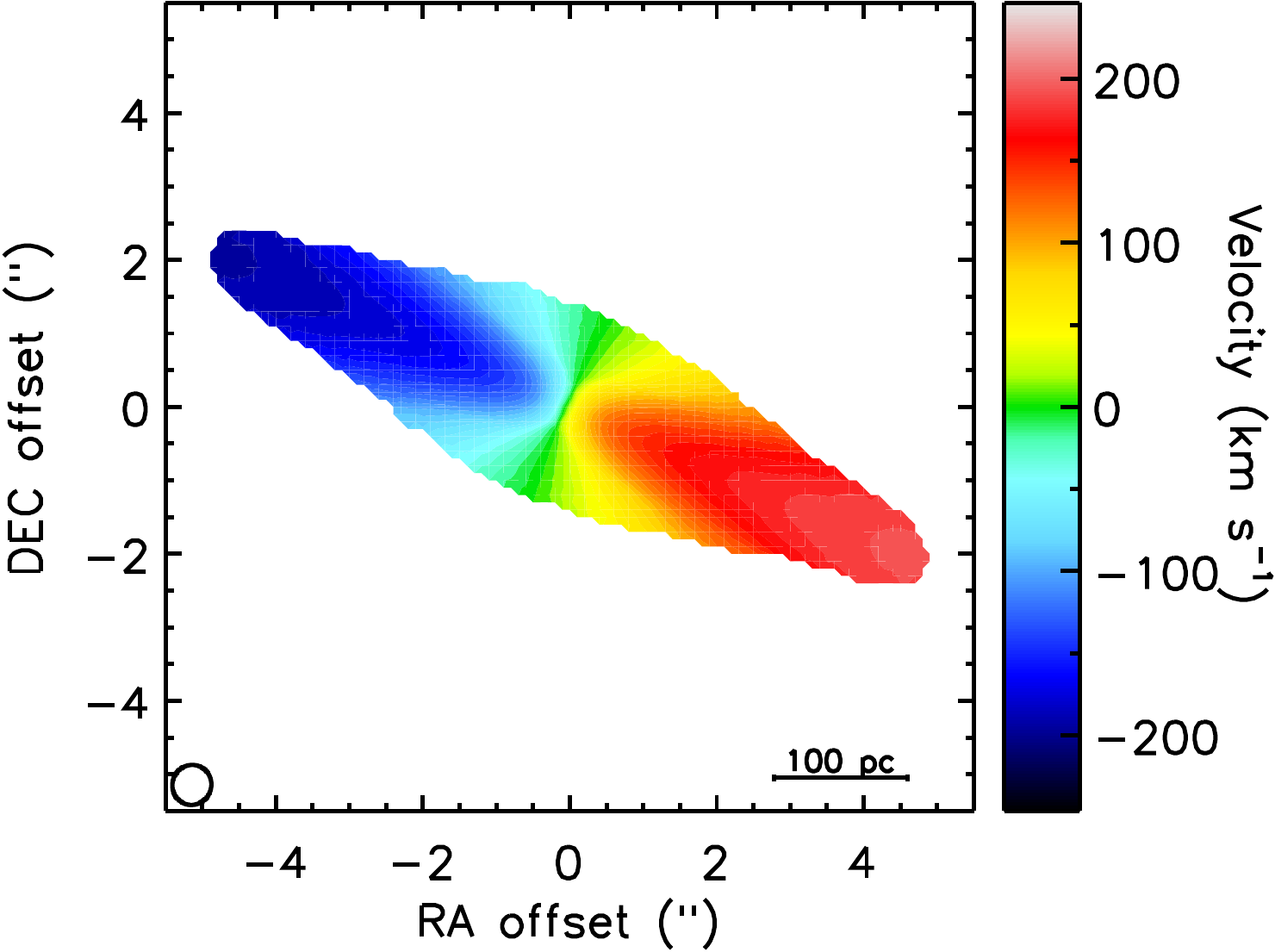}\\
\includegraphics[height=6.9cm,angle=0,clip,trim=-0.6cm 0cm 3.0cm 0.0cm]{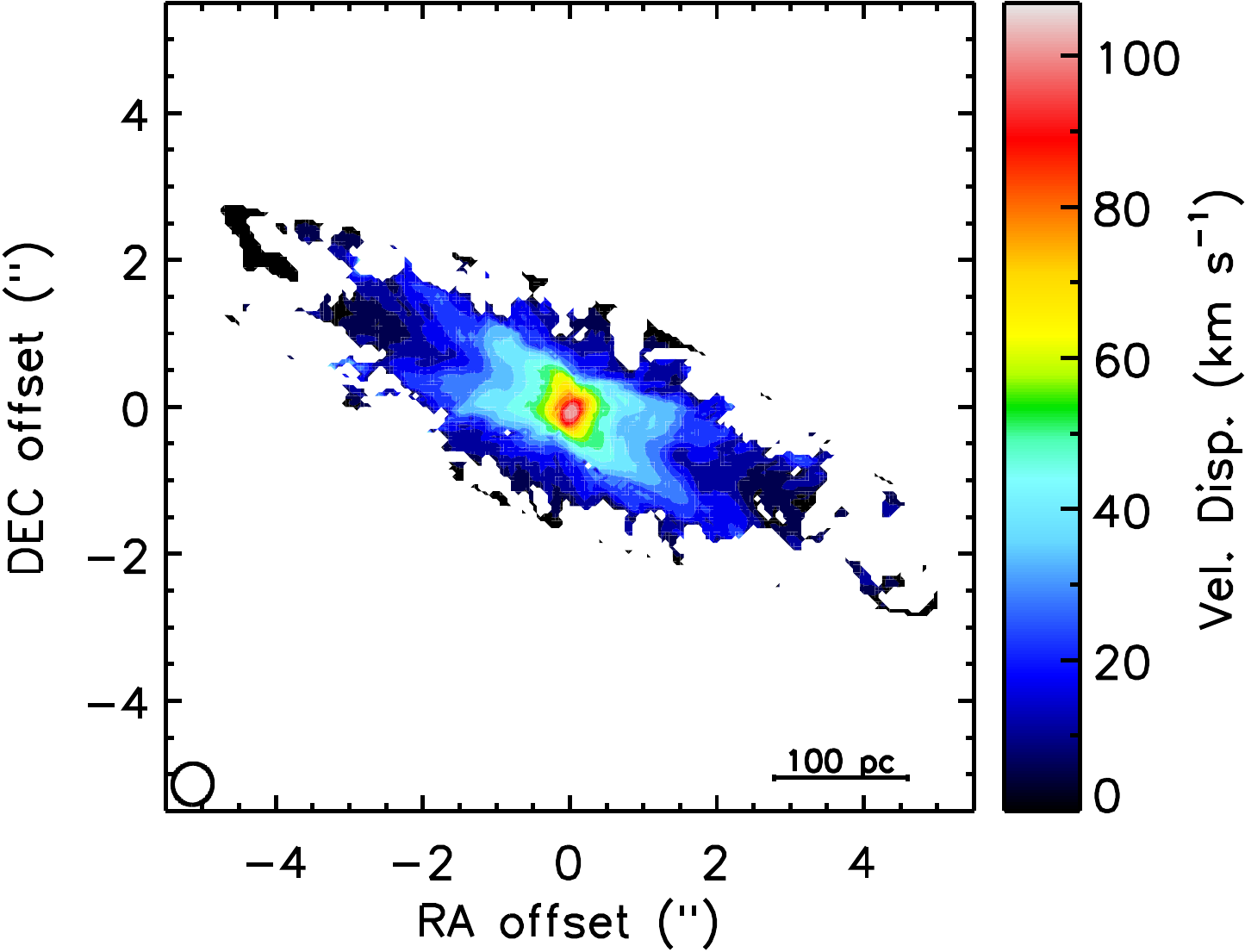}
\includegraphics[height=6.9cm,angle=0,clip,trim=1.8cm 0cm -0.5cm 0.0cm]{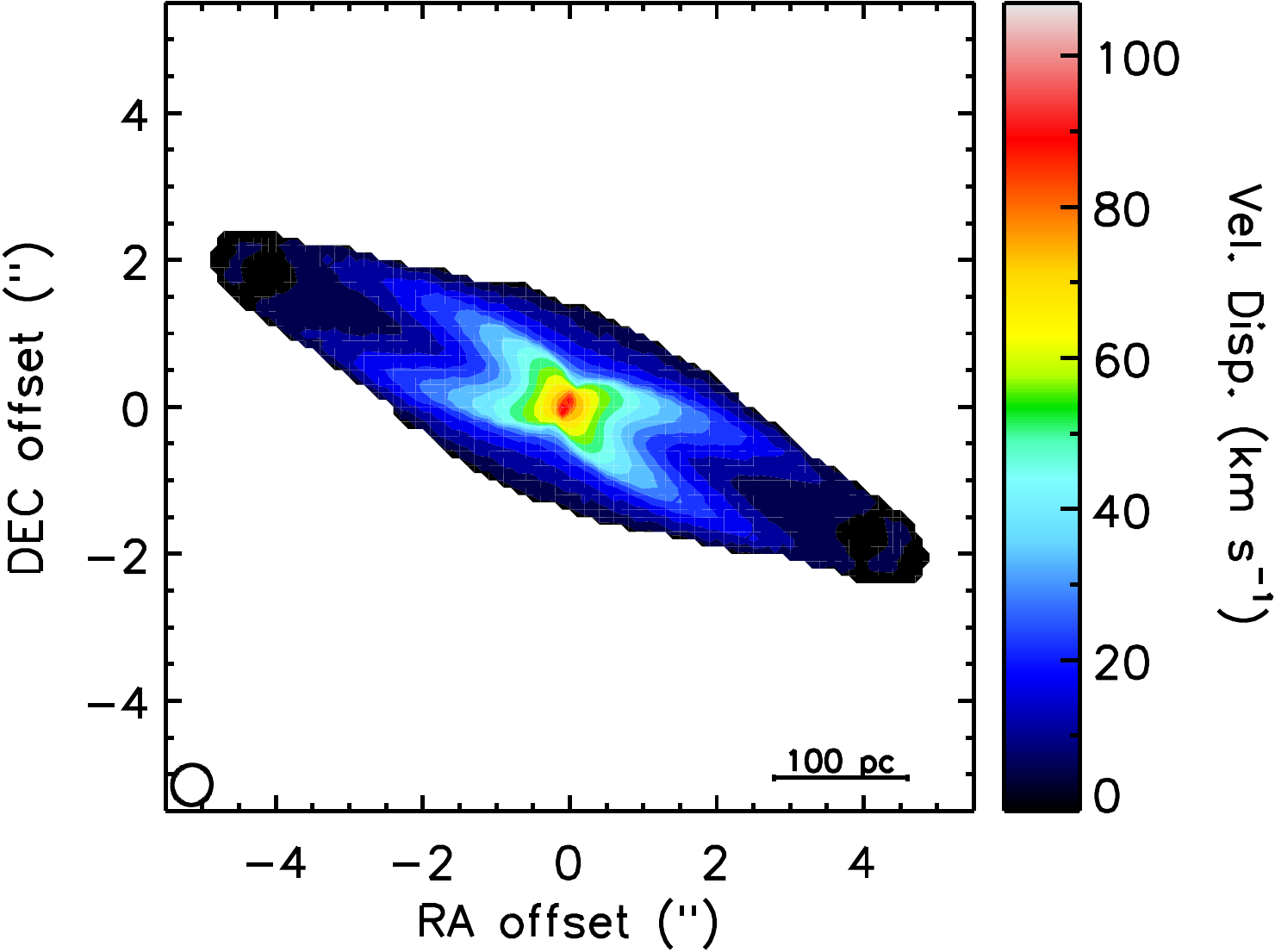}
\caption{Integrated intensity, mean velocity and velocity dispersion maps of the $^{12}$CO(2-1) emission in NGC4697. The moments extracted from the observations are shown in the left panels, while the same moments extracted in an identical way from our best-fit model are shown in the right panels.}
\label{fitpar_compare}
 \end{center}
 \end{figure*}

\begin{figure*} \begin{center}
\includegraphics[height=5.5cm,angle=0,clip,trim=0cm 0cm 0cm 0.0cm]{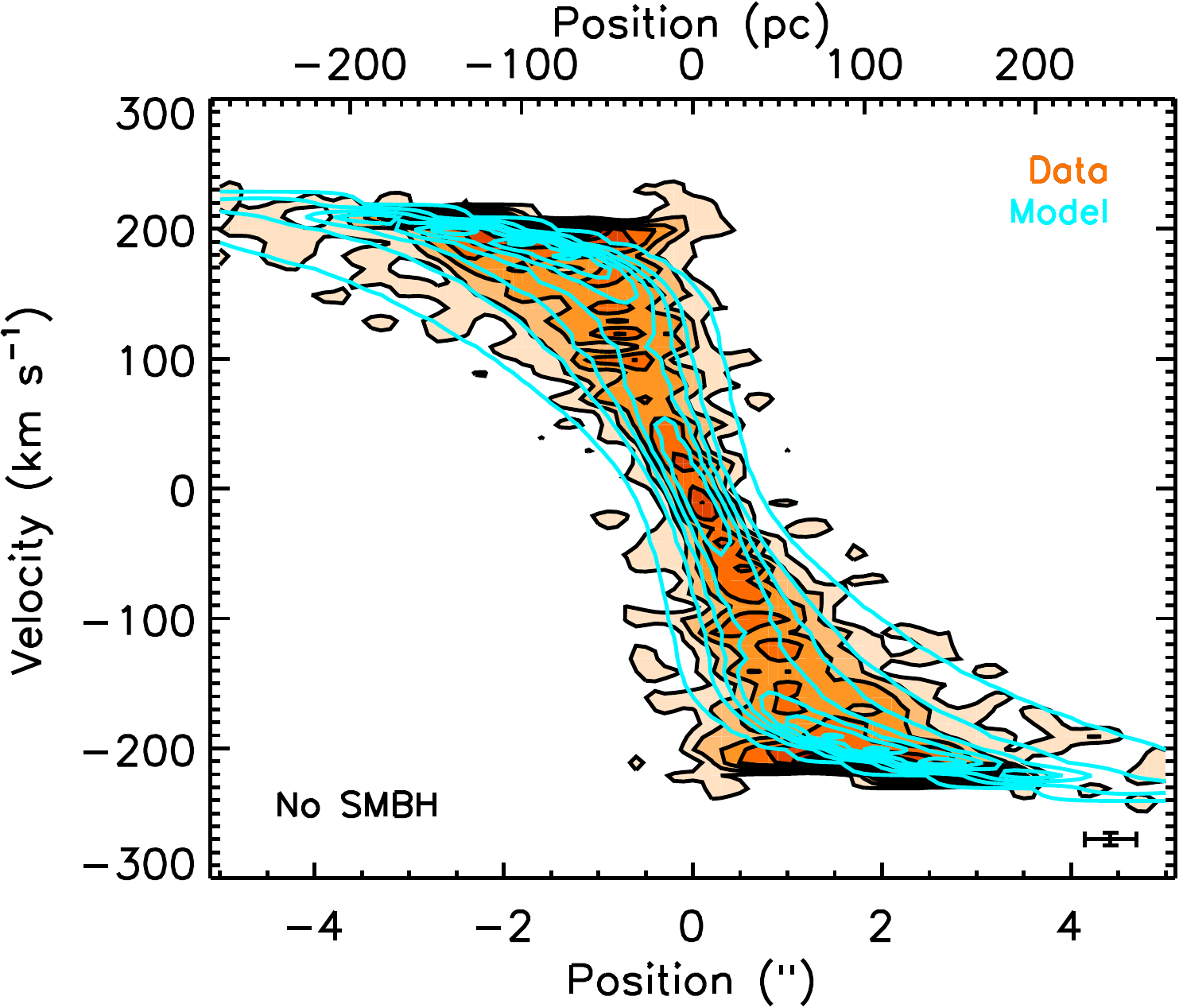}\includegraphics[height=5.5cm,angle=0,clip,trim=2.5cm 0cm 0cm 0.0cm]{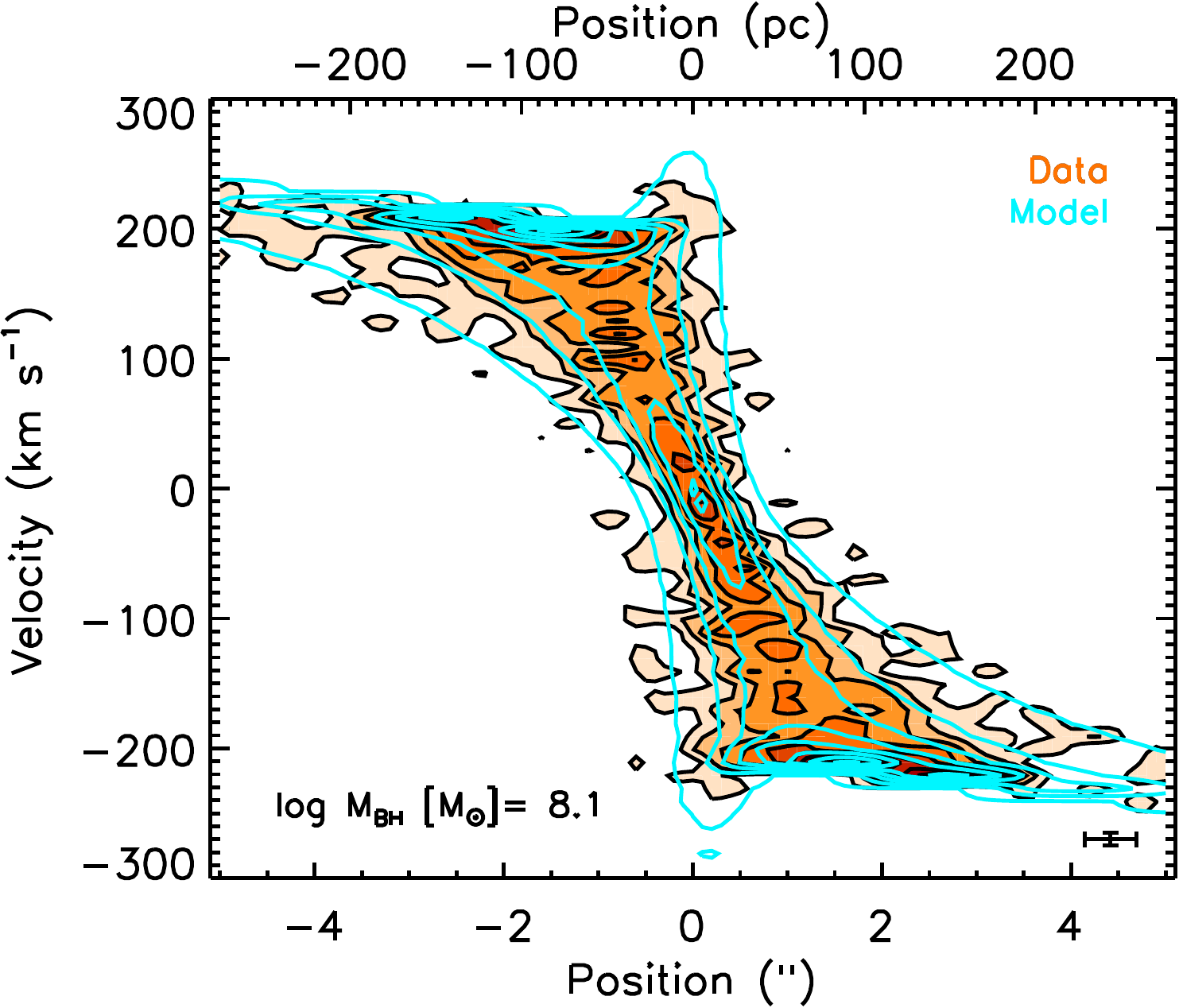}\includegraphics[height=5.5cm,angle=0,clip,trim=2.5cm 0cm 0cm 0.0cm]{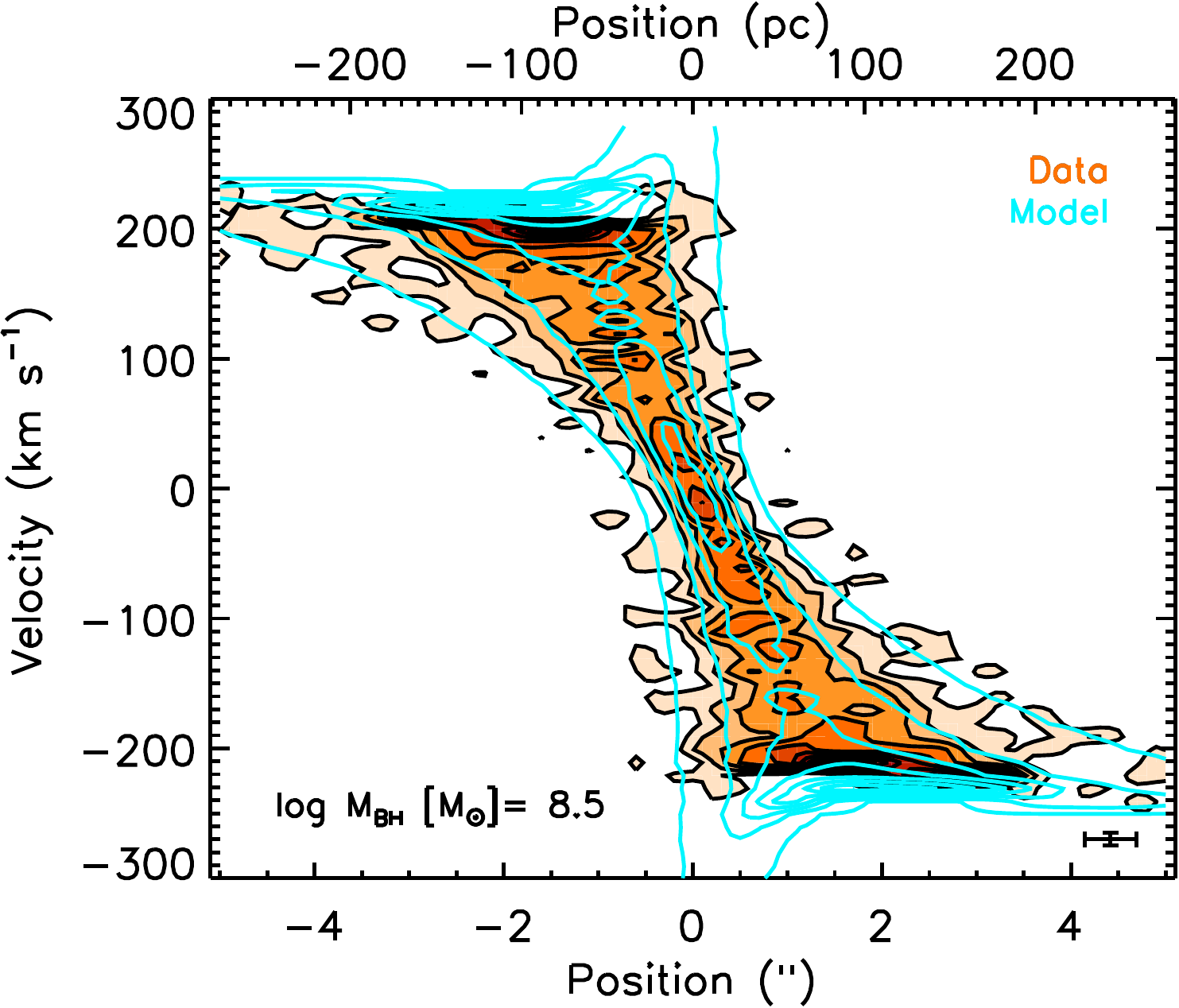}
\caption{As Figure \ref{pvdplot}, but overlaid with model PVDs extracted in an identical fashion from models that only differ by the central SMBH mass (blue contours). The left panel has no SMBH, the centre panel shows our best-fit SMBH mass, and the right panel has an overly massive SMBH. The legend of each panel indicates the exact SMBH mass used. A model with no SMBH is clearly not a good fit to the data. }
\label{threepvdplot}
 \end{center}
 \end{figure*}

We clearly detect the presence of a massive dark object in the centre of NGC4697, with a mass of (1.3$_{-0.14}^{+0.18}$) $\times$10$^8$ \msun (at the 99\% CL). The best-fit $i$-band mass-to-light ratio is 2.14$_{-0.05}^{+0.04}$ \msun/L$_{\odot,i}$. We note that the uncertainties quoted here include random errors only, but we discuss systematic sources of uncertainty in Section \ref{discuss}. The best model is an excellent fit ($\chi^2_{\rm red}$\,=\,1.02). Figure \ref{fitpar_compare} compares the moments extracted from our best-fit \textsc{KinMS} model to those observed.  Figure \ref{threepvdplot} shows the observed PVD over-plotted with the best-fit model, and with models with no SMBH and an overly large SMBH. A SMBH is clearly required to match the Keplerian uptick in velocity around the centre of the galaxy, and the majority of the structure shown in the data can be reproduced by our simple model.

\subsection{Main uncertainties}
\label{uncertainties}
In this Section we consider the main uncertainties that could affect the results derived above. These can roughly be divided into two subsets: issues with the modelling process, and breakdown of the simplifying assumptions made.

\subsubsection{Luminous mass models, AGN \& dust}

We use MGE models of the stellar light distribution to model the contribution of luminous matter to the observed gas kinematics.
However, it is possible for these models to be contaminated, which would lead to an under/overestimate of the stellar luminosity in some parts of our object, and thus bias the derived parameters.

The first of these contaminants is optically obscuring dust, that is ubiquitously present in ETGs with molecular gas \citep{2013MNRAS.432.1796A}.
To minimise dust extinction, our MGE models were constructed from \textit{HST} F814LP images (the longest wavelength available). The models are also carefully fitted to avoid any contamination from extinction still visible in the image (via masking of affected regions). This has been shown to work well to recover the intrinsic light distribution of such systems \citep{2002MNRAS.333..400C}. In addition, we are helped by the fact that the very centre of the galaxy appears relatively extinction free. Dust is present in the outer parts of the molecular disc, however, where the $M$/$L$ is primarily constrained. If anything, we thus expect that residual dust extinction would cause our dynamical $M$/$L$ to be biased to high values, and in turn the SMBH mass to be underestimated.

A second potential issue is that we have assumed that the $M$/$L$ of the stellar population is constant within our field of view. In objects with molecular gas and star formation this may not, however, be strictly correct. \cite{2016arXiv160903559D} addressed this issue in some detail and found that, if present, $M$/$L$ gradients in molecular gas-hosting ETGs like NGC4697 are generally very shallow, due to the low mass fraction of young stars. This is especially likely to be the case in NGC4697 given its very low SFR. This suggests any effect from a variable $M$/$L$ will be small. If any effect is present, then the SMBH mass we derive will in any case be a lower limit (as any star formation would lead to a $M$/$L$ in the galaxy centre lower than we have estimated here, so a larger SMBH would be required to reproduce the same kinematics).

Despite the care taken, the possibility remains that the observational uncertainties and systematics discussed above affect the measurement of our parameters. In addition, there are uncertanties inherent to the construction of the MGE model itself. 
To test the sensitivity of our result to such errors, we re-ran our fitting procedure assuming our MGE model over/underestimated the circular velocity at each radius by $\pm$10\%. In reality such errors would be expected to affect the rotation curve differently at different radii, but we have no a-priori way of knowing which variations to test, and a Monte-Carlo approach is prohibitively time consuming. This approach thus attempts to show the maximum expected variation, and yields a simple estimate of the sensitivity of our fit to such errors. The value of $\pm$10\% is chosen arbitrarily, but it does reflect the typical errors assumed in MGE modelling by other authors (see e.g. \citealt{2013Natur.494..328D}). 
  
 As expected, forcing a variation in the  circular velocity by $\pm$10\% changed our best-fit $M$/$L$ significantly, to 1.7 and 2.6 \msun/L$_{\odot,i}$. Despite these large changes in $M$/$L$, the SMBH mass we derive remains constant at (1.3$_{-0.14}^{+0.18}$)$\times$10$^8$ \msun. This is because we resolve the Keplerian velocity increase in the galaxy centre, that provides $M$/$L$-independent information on the SMBH mass. We thus consider our SMBH mass estimate robust to reasonable systematic errors in our mass models (including $M$/$L$ gradients and dust extinction). 
 
 All SMBH mass estimates are systematically affected by the distance they assume to their target galaxy (with $M_{\rm BH} \propto D$). Here we used an SBF distance from \cite{Tonry:2001ei}, that has a formal uncertainty of $\approx$10\%. Thus the systematic distance uncertainties on the SMBH mass are of a similar order as our random uncertainties. 
  
A further possible contaminant in our target is the AGN. NGC4697 does have some central radio emission and a nuclear hard X-ray source that suggest ongoing low-level accretion. A nuclear point source is also found in our optical image. As mentioned above, we attempted to remove the contribution of the AGN to the light by removing the central spatially unresolved Gaussian from our MGE model, mimicking point-spread function (PSF) subtraction. However, including this point source in the stellar light model would only yield a slightly lower SMBH mass of (9.3$_{-1.4}^{+1.8}$)$\times$10$^7$ \msun. We include this additional 0.1 dex contribution to the SMBH mass error budget, which becomes (1.3$_{-0.17}^{+0.18}$)$\times$10$^8$ \msun.

\subsubsection{Non-circular motions}
As previously discussed, in our analysis we assume that the gas in NGC4697 is in purely circular motion. If significant non-circular motions are present (e.g. inflow, outflow, streaming), then this could affect our analysis.

\cite{2015MNRAS.454.3743R} studied the effect of non-circular motions on the derivation of mass profiles in detail, and showed that dramatic variations can be caused in strongly barred galaxies if the bar is orientated at specific angles with respect to our line of sight. However, NGC4697 is unbarred. Non-circular motions can still be present in objects without bars, for instance because of other departures from axisymmetry in the gravitational potential (e.g. spiral arms), but such motions are low in amplitude compared to the rotation of these systems (e.g. $\approx$10 \kms\ in M51, a galaxy with much stronger spiral structure than early-types; \citealt{2013ApJ...779...45M}). 
In addition, our \textit{HST} images show no sign of non-axisymmetric structures in the stellar mass of this galaxy, suggesting that any such structure would have to be purely gaseous (and thus weak). Furthermore, the best-fit model we present here does not show any major departure when compared to the data (see Fig. \ref{fitpar_compare} and \ref{fitpar_compare_chanmap}), suggesting that there is no significant non-circular motion within the disc.

\subsubsection{Gas velocity dispersion}

In addition to non-circular motions, the molecular gas velocity dispersion provides an additional source of uncertainty \citep[see][]{2016ApJ...822L..28B}. In the models presented here we allow for a single characteristic velocity dispersion within the disc. 
The uncertainty in this measurement is marginalised over when estimating the confidence limits of the other parameters. 
However, as mentioned in Section  \ref{data}, our best-fit value for the velocity dispersion is $\sigma_{\rm gas} \approx$ 0.8 \kms, not well constrained given our channel width of 10 \kms. 
The ability to constrain the velocity dispersion from interferometric observations depends both on the channel width and on the signal-to-noise ratio ($S$/$N$) of the line detection. In the case of a moderate $S$/$N$, the smallest dispersion one can constrain is approximately $2\sqrt{2\ln{2}}$ times smaller than the channel width, or 4.2 \kms\ for our data.

To investigate what the true gas velocity dispersion is, we re-reduced our ALMA data with a channel width of 3~\kms. This width is approximately twice the raw spectral resolution, the smallest size usable if one wishes to avoid correlations between pixels in neighbouring channels (due to the Hanning smoothing applied by the ALMA pipeline).  
At this velocity resolution, we can constrain $\sigma_{\rm gas}$ if it is larger than 1.3 \kms. Emission is still clearly detected in this cube with an RMS of 2.1 mJy beam$^{-1}$. The clear channelisation of the velocity field is also still present, suggesting that the velocity dispersion truly is very low. 

We repeated the kinematic fitting described above (see Section \ref{fitting}) on this higher velocity resolution cube. These fits were able to better constrain the velocity dispersion, yielding a value of 1.65$^{+0.68}_{-0.65}$ \kms\ (at the 99\% CL). We note that adopting this slightly higher value does not affect the results presented in Table \ref{fittable}. We nevertheless discuss this surprisingly low velocity dispersion further in Section \ref{veldispdiscuss}.

In all the analyses above we assumed that the gas velocity dispersion was spatially constant. In reality the velocity dispersion could vary with radius and azimuth within the gas disc. In the central part of the galaxy, where beam smearing is important, an increase in velocity dispersion could lead to a SMBH mass overestimate.

To quantify the size of this effect we re-ran the modelling process for NGC4697, allowing for a variable velocity dispersion as a function of radius. {When we allow for a linear gradient in velocity dispersion, we find the same low velocity dispersion values and a marginally negative slope, consistent with zero.}
To completely remove the SMBH kinematic signature and simultaneously reproduce our data, the velocity dispersion of the gas would have to increase by a factor of 50, from $<$2 \kms\ outside a galactocentric radius of $\approx$50~pc to over 130~\kms\ inside our inner resolution element. We consider this highly unlikely. 

While the velocity dispersions of clouds in the centre of the Milky Way are larger than those of clouds in the disc, and several individual colliding clouds do have large line-widths, when averaged by volume one would typically expect an enhancement of only a factor of $\approx$2 \citep[e.g.][]{1997MNRAS.292..871K}. Even a factor of 20 increase over our measured disc average velocity dispersion would have a negligible effect on our measured parameters. We thus consider them robust again such a change.

\subsubsection{Gas mass}

In addition to luminous matter, the mass of any ISM material present also contributes to the total dynamical mass of a galaxy.
As we measure the CO line in this work, we could in principle include the molecular gas mass density in our {mass models}. However, given the uncertainty in $X_{\rm CO}$ (the conversion between $^{12}$CO(1-0) luminosity and mass) and the $^{12}$CO(1-0)/$^{12}$CO(2-1) ratio, this would add additional uncertainty, {especially in objects with high gas fractions}.

Luckily, due to their low gas fractions, in ETGs the molecular reservoirs are usually dynamically unimportant (see e.g. \citealt{2016arXiv160903559D}).  This is certainly true here, where the total H$_2$ mass over the entire molecular gas disc is only $\approx10^7$ \msun\ (assuming a standard Galactic conversion factor). {Our mass model suggests a stellar mass of $\approx$2$\times$10$^9$ \msun\ within 200\,pc of the centre of NGC4697, two orders of magnitude greater than the molecular gas mass in the same region. Additionally, the total molecular gas mass is only a tenth of the SMBH mass}. The H$_2$ mass within our inner resolution element (i.e. around the SMBH) is even less (a few times 10$^6$ \msun\ of molecular gas is seen in projection against the nucleus). We note that no information is available on the \hi\ mass of NGC4697, but we consider it unlikely that this substantially biases our results, as the inner parts of massive ETGs are always molecular gas-dominated when both phases are present \citep{2012MNRAS.422.1835S,2014MNRAS.444.3427D}.
 We thus consider it unlikely that neglecting the mass of gas in our fitting procedure significantly biases our results.

\begin{figure} \begin{center}
\includegraphics[width=0.48\textwidth,angle=0,clip,trim=0cm 0cm 0cm 0.0cm]{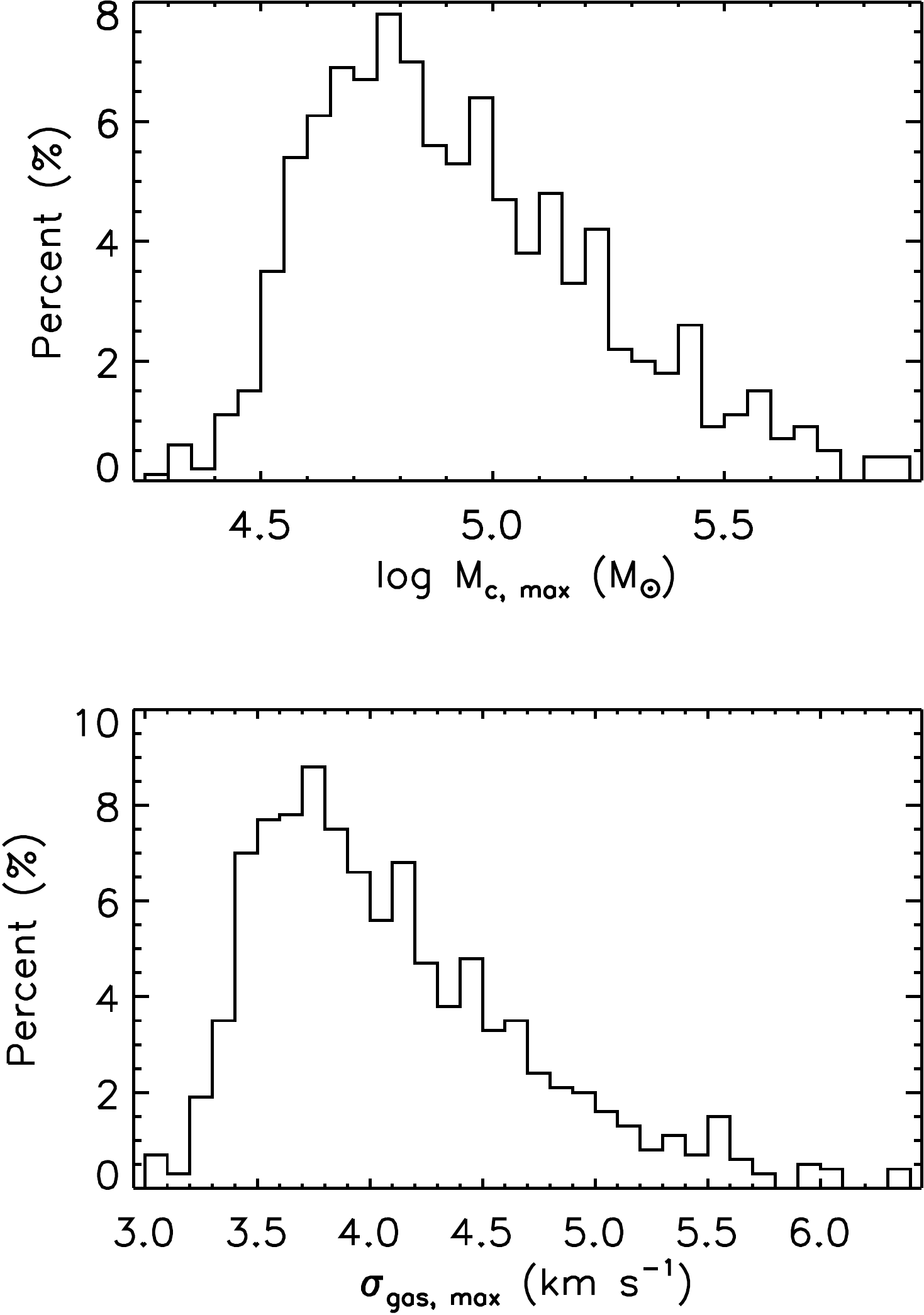}
\caption{\textit{Top panel:} Likelihood of a given GMC being the largest hosted by NGC4697 as a function of mass. \textit{Bottom panel:} As above, but as a function of the cloud velocity dispersion. These maximum masses and velocity dispersions are higher than found in our observations, suggesting that other causes of the low velocity dispersion are required.}
\label{gmc_massfunc}
 \end{center}
 \end{figure}

\section{\uppercase{Discussion}}
 \label{discuss}
 
 In this paper we presented ALMA $^{12}$CO(2-1) observations of the nearby fast-rotator NGC4697. We showed that it hosts a small molecular gas disc, cospatial with a dust disc that we also mapped in thermal dust emission. We then used the observed gas kinematics to estimate the mass of the central SMBH, along with other galaxy parameters.
 In this Section we compare our results to those of other authors and discuss several additional issues.
 
\subsection{Molecular gas mass}

In Section \ref{data} we estimated that the molecular gas mass present in NGC4697 is (1.62$\pm$0.17)$\times$10$^7$ \msun.
This object has not been interferometrically mapped before, but it was observed with the single-dish Institut de Radioastronomie Millim\'etrique (IRAM) 30m telescope by \cite{2011MNRAS.414..940Y}. 
These authors did not detect $^{12}$CO(1-0) or $^{12}$CO(2-1) in NGC4697, and set a 3$\sigma$ upper limit of (7$\pm$2.4)$\times$10$^6$ \msun\ of molecular gas (using the same standard $X_{\rm CO}$ and distance as we do in Section \ref{data}). 
However, they assumed a velocity width of 300 \kms, while here we detect emission over $\approx$480 \kms. Correcting the estimate of \cite{2011MNRAS.414..940Y} for this larger line width yields an upper limit of (1.2$\pm$0.4)$\times$10$^7$ \msun\ of molecular gas, consistent within the errors with the value estimated from our ALMA data.

 \subsection{Gas velocity dispersion}
 \label{veldispdiscuss}
 
 As discussed at some length above, the velocity dispersion in the molecular gas disc of NGC4697 seems abnormally low. 
 Molecular gas velocity dispersions in the Milky Way and nearby spiral galaxies are typically $\approx$6-12 \kms\ \citep[e.g.][]{1981MNRAS.194..809L,2013AJ....146..150C}.  In NGC4697 we find a value of 1.65$^{+0.68}_{-0.65}$ \kms. While there is some uncertainty in this estimate, $\sigma_{\rm gas}$ is certainly $<$3 \kms, {and it does not seem to vary significantly radially}.  
 It is thus important to establish what may cause NGC4697 to differ and have such dynamically cold gas.
 
 Observational effects are one possibility. For instance, \cite{2013AJ....146..150C} and \cite{2016AJ....151...34C} show that interferometric observations can underestimate the velocity dispersion if they resolve out a smooth component of the molecular emission. We do not expect this to be the case here, however, as our observations are sensitive to emission $\ltsimeq$300 pc, which is the total size of the dust disc visible in \textit{HST}, and much larger than the characteristic size of giant molecular clouds (GMCs). In addition, the single-dish upper limit of \cite{2011MNRAS.414..940Y} does not allow for the presence of much additional mass not detected by ALMA.
 
 In normal spiral galaxies, it is thought that the velocity dispersion is set by a feedback loop between gravitational collapse and energy injection from star formation, in such a way that clouds stay approximately in virial equilibrium \cite[e.g.][]{1981MNRAS.194..809L,1987ApJ...319..730S}.  One might speculate that with its low estimated SFR, NGC4697 is simply unable to heat its molecular gas effectively. This is unlikely to be the full story, however, because although the total SFR is low, the SFR surface density within the small nuclear disc is between 0.001 and 0.017 \msun\ yr$^{-1}$ kpc$^{-2}$, the typical range found in spiral galaxy discs \citep{1998ApJ...498..541K}. In addition, a higher SFR would be expected if the gas were collapsing unopposed. 
 
One could also postulate that there are no large GMCs in NGC4697. The \cite{1981MNRAS.194..809L} relations show that larger clouds have higher velocity dispersions:
\begin{equation}
\left(\frac{\sigma_{\rm gas}}{\mathrm{km\,s^{-1}}}\right) =  1.1 \left(\frac{L}{\mathrm{pc}}\right)^{0.38},
\label{larson}
\end{equation}
where $L$ is the linear extent of a cloud.
 GMCs have typical linear extents of 10 to 100 pc, and thus typical velocity dispersions of 2.6 to 6.3 \kms. If the GMC mass function were not fully sampled in NGC4697, then this could help explain the low velocity dispersion. 
This explanation is supported by the morphology of the gas in our observations, as the disc is remarkably smooth even at our high resolution (30 pc).
 The object is fairly edge-on, so projection effects may be important, but given that the majority of the mass (and thus flux) is in the largest GMCs \citep{2005PASP..117.1403R}, some clumpiness would be expected. 
 Given that we do not seem to spatially resolve any clumpy cloud structure with $L$\,>\,30 pc, Equation \ref{larson} predicts that the remaining (smaller) clouds should have a velocity dispersion of $<$4 \kms, and (via another \citealt{1981MNRAS.194..809L} relation) contain $\ltsimeq10^4$ \msun\ of molecular gas each. 
  
This explanation is not fully satisfying, however, as it simply shifts the question to understanding why the GMC mass function is not fully populated in NGC4697.
To address this question, we drew randomly from a typical Galactic GMC mass function (with a power-law slope of -1.75, lower mass cutoff of 100 \msun\ and upper mass cutoff of 10$^6$ \msun; \citealt{2005PASP..117.1403R}) to assemble a total molecular mass of 1.6$\times$10$^7$ \msun. We repeated this process 10,000 times, and show in Figure \ref{gmc_massfunc} histograms of the maximum cloud mass in each realisation and the velocity dispersion expected from this cloud (via Larson's relations; \citealt{1981MNRAS.194..809L}). 
We note that in this analysis we have had to assume that the \cite{1981MNRAS.194..809L} relations hold in NGC4697. Observationally there is some evidence this may not be the case in all environments \citep[e.g.][]{2008ApJ...686..948B,2009ApJ...699.1092H,2015ApJ...803...16U}. Modulo this uncertainty, we find that the median maximum cloud mass expected for a gas disc like that of NGC4697 is 7$\times$10$^4$ \msun, that equates to a velocity dispersion of 3.9 \kms. This suggests that it is quite plausible that the GMC mass function is not fully populated in this object.

On the other hand, all the realisations in our test returned a maximum velocity dispersion $>$3 \kms, while the observed disc in NGC4697 seems to have a velocity dispersion even lower than this. In addition, our analysis neglects any inter-cloud velocity dispersion, that (especially in an edge-on object) should act to increase the measured velocity dispersions. 
Thus, while it is quite likely that the GMC mass function is not fully sampled in NGC4697, we still cannot fully explain the low dispersion observed.

\begin{figure} \begin{center}
\includegraphics[width=0.48\textwidth,angle=0,clip,trim=0cm 0cm 0cm 0.0cm]{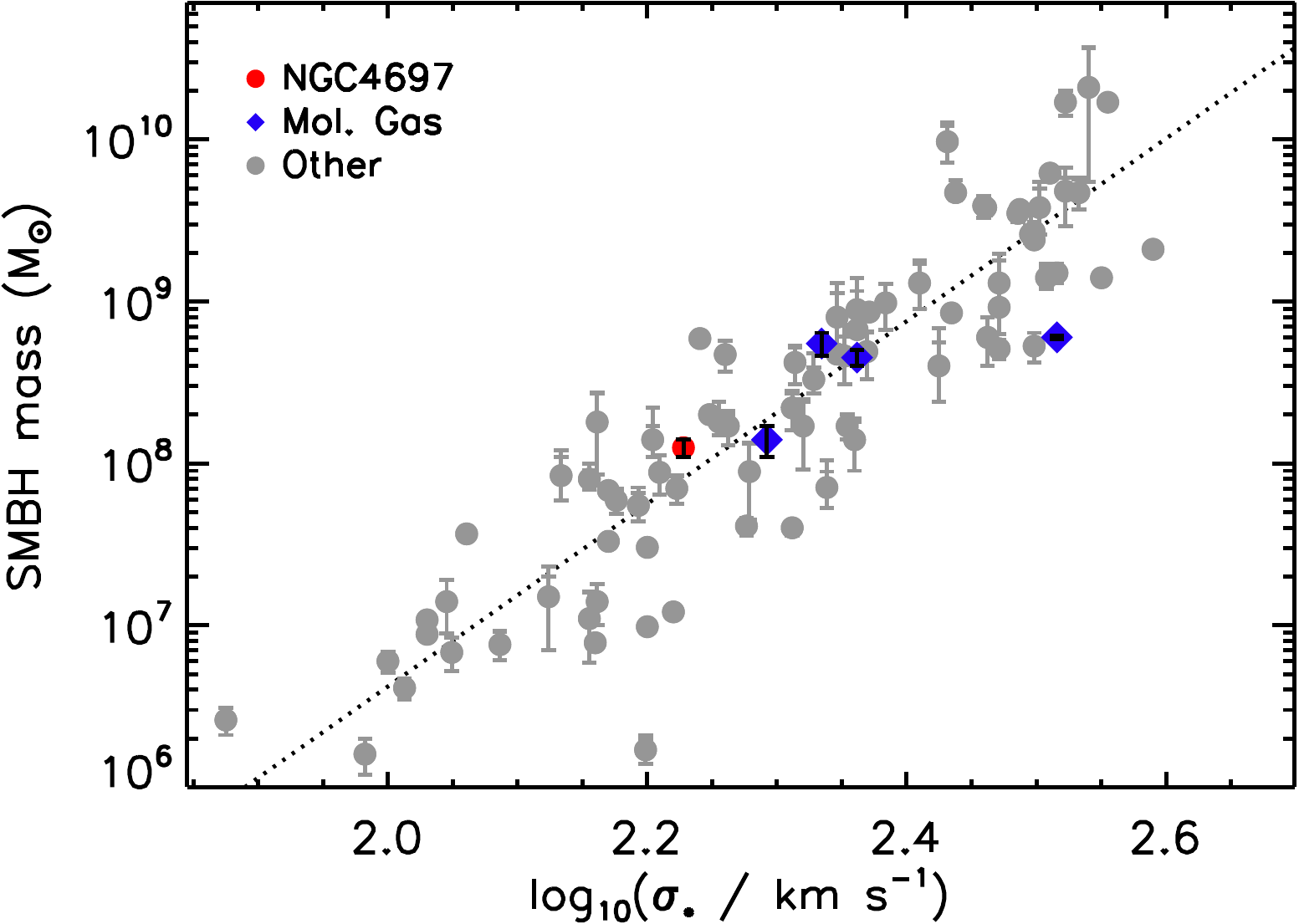}
\caption{$M_{\rm BH}$ -- $\sigma_*$ relation from the compilation of \protect \cite{2013ApJ...764..184M} (grey points and dotted line). We also show the SMBH mass measured for NGC4697 in this paper as a red circle and highlight measurements from other works also using the molecular gas technique with blue diamonds. }
\label{msigmaplot}
 \end{center}
 \end{figure}

Another possible explanation of this discrepancy is the hard radiation field present in NGC4697. This system has a diffuse X-ray halo and many X-ray binaries \citep{2001ApJ...556..533S}, that along with evolved stars could mean the radiation field incident on the cold gas is much harder than that in the Milky Way. In this case, it is possible that the CO molecules are confined deeper inside the molecular clouds, in dynamically colder regions \citep{2011MNRAS.415.3253S,2015MNRAS.452.2057C}. We note that this scenario would also affect $X_{\rm CO}$, leading to a higher total molecular gas mass (and thus an even lower SFE). 
 
Yet another possible explanation is the stabilising influence of the galaxy bulge (so-called `morphological quenching'). Using hydrodynamic simulations \cite{2009ApJ...707..250M,2013MNRAS.432.1914M} showed that the presence of a large bulge can stabilise a low-mass gas disc against star formation, and predicted that such stable discs would have very low velocity dispersions. This could naturally explain the low velocity dispersion of the gas in NGC4697, and its low SFE. Further observations of galaxies with suppressed SFE would be required to confirm which, if any, of these mechanisms can explain this phenomenon. 

  \subsection{Comparison of SMBH mass with other measurements}

In earlier Sections we estimated a mass of (1.3$_{-0.17}^{+0.18}$)$\times$10$^8$ \msun\ for the SMBH in the centre of NGC4697.
As mentioned in the introduction, the SMBH mass in this object has already been measured from observations of the galaxy's stellar kinematics. It is thus instructive to compare these other estimates to ours, to quantify systematics and uncertainties. 

\cite{2003ApJ...583...92G} found $M_{\rm BH}$\,=\,(1.6$\pm$0.2) $\times$10$^8$ \msun\ from Schwarzschild modelling of the stellar kinematics in this source, while \cite{2011ApJ...729...21S} used the same technique but added a model dark matter halo and obtained the same value, (1.6$\pm$0.5) $\times$10$^8$ \msun. These authors both used slightly different distances, that we homogenised to our value here (assuming $M_{\rm BH} \propto D$).
  
  The value we have derived is in excellent agreement with these literature measurements. This is despite the very different systematics involved. 
 For instance, \cite{2003ApJ...583...92G} and \cite{2011ApJ...729...21S} used Schwarzschild modelling of the stellar kinematics to estimate the SMBH mass. This method is very different, both conceptually and numerically, from the technique used here. In addition, both these authors used \textit{HST} F555W observations to construct their mass models, while we used F850LP. They also only had access to long slit spectroscopy when modelling the stellar kinematics, while we had full three-dimensional data. 
Overall this agreement between different methods, using very different tracers, is highly encouraging for the field.

   \subsection{SMBH mass - galaxy correlations}

Individual estimates of SMBH masses are interesting, but it is their variations with galaxy properties that motivate the WISDOM project. In Figure \ref{msigmaplot}, we show the location of our object on the $M_{\rm BH}$ -- stellar velocity dispersion ($\sigma_*$) relation, using the compiled SMBH masses from \cite{2013ApJ...764..184M}.  NGC4697 is in red, and all of the other recent SMBH mass measurements using molecular gas are in blue (\citealt{2013Natur.494..328D,2015ApJ...806...39O,2016ApJ...822L..28B}; Onishi et al., 2017). NGC4697 lies slightly above the best-fit $M_{\rm BH}$ -- $\sigma_*$ relation of \cite{2013ApJ...764..184M}, but well within the scatter.

NGC4697 is only the fifth object studied to date using molecular gas, and it has the lowest SMBH mass studied thus far. Four of these objects seem to cluster tightly around the best fit $M_{\rm BH}$ -- $\sigma_*$ relation of  \cite{2013ApJ...764..184M}, while the fifth (NGC1332) falls below, at the outer edge of the scatter.  It is hard to make strong statements about the importance of this without better statistics. We note, however, that the relatively small errors of these measurements are highly promising for studies of the intrinsic scatter in the SMBH -- galaxy relations.

\section{\uppercase{Conclusions}}
\label{conclude}

 In this paper we have presented ALMA $^{12}$CO(2-1) observations of the nearby fast-rotating ETG NGC4697, taken as part of the WISDOM project. 
  This galaxy hosts a small central molecular gas disc, co-spatial with an obscuring dust disc visible in \textit{HST} imaging, and containing (1.62\,$\pm$\,0.01\,$\pm$\,0.36)\,$\times$\,10$^7$ \msun\ of molecular hydrogen assuming a Galactic $X_{\rm CO}$ factor.
 We also detected spatially-resolved 1 mm continuum emission from this disc, that seems to be dominated by the Rayleigh-Jeans tail of the thermal emission from dust. We used this emission, along with data from the literature, to estimate a dust mass of (2.8\,$\pm$\,0.2\,$\pm$\,1.0)$\times10^5$ \msun, a dust temperature of 28.7$\pm$0.4 K, and a molecular gas-to-dust ratio of 58$\pm$7.

 The position-velocity diagram extracted along the major axis NGC4697 shows a Keplerian increase of the rotation velocity inside the sphere of influence of the central SMBH. 
A forward modelling approach in a Bayesian framework was used to fit the observed data cube of NG4697 and estimate the SMBH mass, stellar $M$/$L$, and numerous parameters describing the structure of the molecular gas disc.
We found that the SMBH in this galaxy has a mass of (1.3$_{-0.17}^{+0.18}$) $\times$10$^8$ \msun\ (at the 99\% CL) and the $i$-band mass-to-light ratio is 2.14$_{-0.05}^{+0.04}$ \msun/L$_{\odot}$. The inclination of the system is constrained to be 76$\pm$1$.\!\!^{\circ}$1.
With this SMBH mass, NGC4697 lies slightly above the best-fit $M_{\rm BH}$ -- $\sigma_*$ relation of \cite{2013ApJ...764..184M}, but well within the scatter.

NGC4697 was found to have a very low molecular gas velocity dispersion. Part of the explanation is likely that the GMC mass function is not fully sampled in objects with such low H$_2$ masses. This interpretation is supported by the lack of spatially-resolved structure in the integrated intensity map of NGC4697. However, other physical mechanisms are probably required to explain the low dispersion in this object. 
 It is possible that CO molecules are confined deeper inside the molecular clouds by a hard radiation field, or that the large bulge of this system stabilises the gas disc (as expected from simulations of morphological quenching).

As a technique in its infancy, it is important to cross-check the results of molecular gas SMBH mass estimates with those made using other techniques. For NGC4697, we find that our estimate of the SMBH mass is entirely consistent with previous measurements using stellar kinematics. This is despite these studies using a different technique and completely different data.
{This is in contrast to NGC1332, where \cite{2016ApJ...822L..28B, 2016ApJ...823...51B} used molecular gas kinematics to find an SMBH mass significantly lower than that estimated using stellar kinematics. 
Larger samples with SMBH masses derived from both molecular gas dynamics and stellar dynamics are clearly required to understand this discrepancy.  In addition, cross-checks with other techniques such as ionised gas and maser SMBH mass measurements will be important. In this way, we can build on the promising results of this work and allow molecular gas SMBH mass estimates to be used with confidence in the ALMA era.}

 \vspace{0.5cm}
\noindent \textbf{Acknowledgments}

TAD acknowledges support from a Science and Technology Facilities Council Ernest Rutherford Fellowship, and thanks F. van de Voort for useful discussions and comments on early versions of this manuscript.
MB was supported by the consolidated grants `Astrophysics at Oxford' ST/H002456/1 and ST/K00106X/1 from the UK Research Council.
MC acknowledges support from a Royal Society University Research Fellowship.

This paper makes use of the following ALMA data: ADS/JAO.ALMA\#2015.1.00598.S. ALMA is a partnership of ESO (representing its member states), NSF (USA) and NINS (Japan), together with NRC (Canada), NSC and ASIAA (Taiwan) and KASI (Republic of Korea), in cooperation with the Republic of Chile. The Joint ALMA Observatory is operated by ESO, AUI/NRAO and NAOJ.

This paper also makes use of observations made with the NASA/ESA Hubble Space Telescope, and obtained from the Hubble Legacy Archive, which is a collaboration between the Space Telescope Science Institute (STScI/NASA), the Space Telescope European Coordinating Facility (ST-ECF/ESA) and the Canadian Astronomy Data Centre (CADC/NRC/CSA). This research has made use of the NASA/IPAC Extragalactic Database (NED) which is operated by the Jet Propulsion Laboratory, California Institute of Technology, under contract with the National Aeronautics and Space Administration.

\bsp
\bibliographystyle{mnras}
\bibliography{bibNGC4697.bib}
\bibdata{bibNGC4697.bib}
\bibstyle{mnras}

\label{lastpage}

\appendix

\section{Channel maps}
\label{channelmaps}

\begin{landscape}

\begin{figure} \begin{center}
\includegraphics[width=24cm,angle=0,clip,trim=0cm 0cm 0cm 0.0cm]{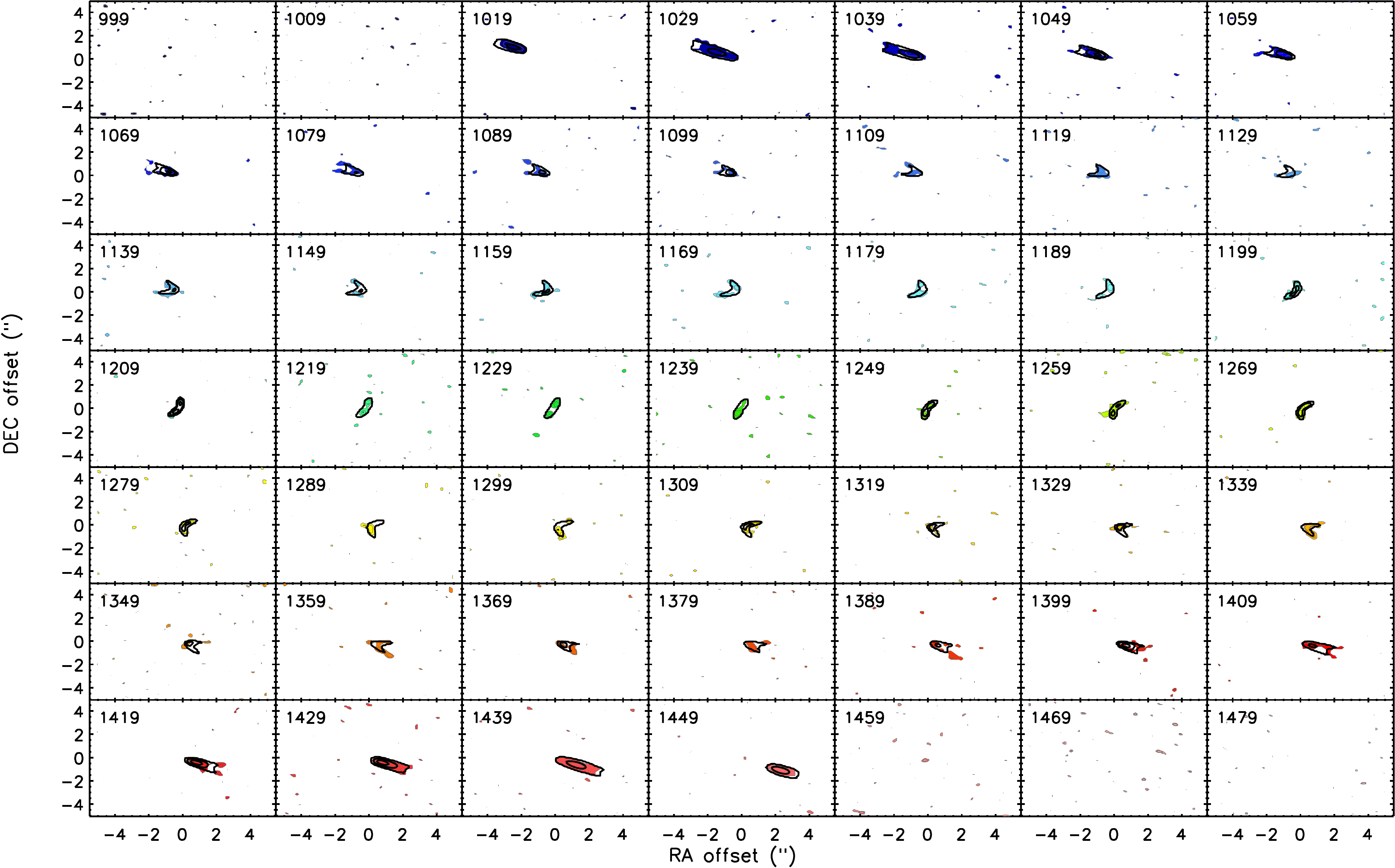}
\caption{Channel maps of our ALMA $^{12}$CO(2-1) data in the velocity range where emission is detected (1019-1449 \kms). The legend at the top left of each panel shows the velocity of that channel (in \kms). The coloured regions with grey contours shows the areas detected with more than a 2.5$\sigma$ significance. Overplotted on this in black are the same contour levels from our best-fit model. Our model agrees well with the observed data in every velocity channel.}
\label{fitpar_compare_chanmap}
 \end{center}
 \end{figure}

\end{landscape}

\end{document}